\documentclass[11pt]{article}
\usepackage{color, amsmath, amssymb, amsthm, graphicx, bbm}
\usepackage[top=1in, bottom=1in, left=1in, right=1in]{geometry}

\usepackage[toc,page]{appendix}
\usepackage{tikz}
\usepackage{titling}
\allowdisplaybreaks

\usepackage{setspace}
\usepackage{adjustbox}
\usepackage[colorlinks,allcolors=blue]{hyperref}
\usepackage[round]{natbib}  

\usepackage{color}
\usepackage{caption}
\usepackage{amsfonts}
\usepackage{graphicx}
\usepackage{mathrsfs}
\usepackage{booktabs}
\usepackage[flushleft]{threeparttable}
\usepackage{xr}
\usepackage{amsmath}
\usepackage{amssymb}
\usepackage{float}
\usepackage{verbatim}

\renewcommand{\baselinestretch}{1.4}
\usepackage{geometry}
\makeatletter 
\def\singlespace{\def\baselinestretch{1}\@normalsize}

\@addtoreset{equation}{section}

\numberwithin{equation}{section}
\theoremstyle{plain}

\newtheorem{theorem}{Theorem}

\newtheorem{corollary}{Corollary}

\newtheorem{proposition}{Proposition}

\newtheorem{assumption}{Assumption}

\def\remark#1{\noindent{\bf Remark #1\ }}

\def\FDP{\mathrm{FDP}}

\def\cov{\mathrm{cov}}
\def\var{\mathrm{var}}

\def\z{\tilde{z}}
\def\Card{\mathrm{Card}}
\def\ID{\mathbb{I}}

\def\X{{\bf X}}
\def\mC{{\mathcal C}}

\def \bX{{\mathbf{X} }}

\def\bbe{{\bm{\beta}}}
\def\sT{{\top}}

\newcommand{\bs}{\boldsymbol}
\newcommand{\bm}{\boldsymbol}

\def\hat{\widehat}
\def\tilde{\widetilde}

\def \APSE{\mathrm{APSE}}

\def\E{\mathbb{E}}
\def\bI{\mathbb{I}}

\title{Dynamic statistical inference in massive datastreams}
\author{Jingshen Wang\thanks{Division of Biostatistics, UC Berkeley} \and Lilun Du \thanks{School of Business and Management, Hong Kong University of Science and Technology} \and Changliang Zou \thanks{School of Statistics and Data Sciences, Nankai University} \and Zhenke Wu \thanks{Department of Biostatistics, University of Michigan, Ann Arbor} }

\date{\today}

\begin{document}

	\maketitle

\begin{abstract}
	Modern technological advances have expanded the scope of applications requiring analysis of large-scale datastreams that comprise multiple indefinitely long time series. There is an acute need for statistical methodologies that perform online inference and continuously revise the model to reflect the current status of the underlying process. In this manuscript, we propose a dynamic statistical inference framework--named  dynamic tracking and screening (DTS)--that is not only able to  provide accurate estimates of the underlying parameters in a dynamic statistical model, but also capable of rapidly identifying irregular individual streams whose behavioral patterns deviate from the majority. Concretely, by fully exploiting the sequential feature of datastreams, we develop a robust estimation approach under a framework of varying coefficient model. The procedure naturally accommodates unequally-spaced design points and updates the coefficient estimates as new data arrive without the need to store historical data. A data-driven choice of an optimal smoothing parameter is accordingly proposed. Furthermore, we suggest a new multiple testing procedure tailored to the streaming environment. The resulting DTS scheme is able to adapt time-varying structures appropriately, track changes in the underlying models, and hence maintain high accuracy in detecting time periods during which individual streams exhibit irregular behavior. Moreover, we derive rigorous statistical guarantees of the procedure and investigate its finite-sample performance through simulation studies. We demonstrate the proposed methods through a mobile health example to estimate the timings when subjects' sleep and physical activities have unusual influence upon their mood.
\medskip


	\end{abstract}

\section{Introduction}\label{Sec-1}
	\subsection{Background and motivation}

Highly developed information and sensor technologies constantly generate and store massive longitudinal data sets that become available sequentially at a high frequency. Ranging from telecommunications \citep{black03}, environmental monitoring  \citep{Guerriero_etal_2009}, retail banking \citep{Tsung07}, health care \citep{Spiegelhalter_etal_2012}, and network monitoring  \citep{Vaughan_2013}, such a type of data collection is pervasive and is referred to as streaming data throughout this manuscript. Other than the high-frequency feature, as massive datastreams are often collected from distinct classes of subjects often in highly dynamic real-life environments, it is commonly believed that they may contain a growing number of irregular patterns \citep{Gama_2010}.

In this context, a statistical methodology that is relevant to streaming data analysis often pertains to algorithms that enable us to (1) dynamically revise statistical models and update the statistical inferential results by incorporating local dynamic changes, (2) efficiently store summary statistics from past history without the need to store an ever-increasing data history \citep{Aggarwal_2006}, and (3) identify individual datastreams whose behavioral patterns deviate significantly from that of most individuals.
In this manuscript, we propose a dynamic statistical inference procedure--so called {\it dynamic tracking and screening} (DTS)--which is able to temporally adapt to time-varying structures, to incorporate time-varying covariates and to identify irregular datastreams as soon and accurately as possible.

Our motivating example comes from the Intern Health Study (IHS)--an ongoing multi-site cohort study enrolled
more than 3,000 medical interns--which aims to assess behavioral phenotypes that precipitate stress episodes and mood changes during the first year of residency training \citep{kalmbach2018effects, kious2019altitude}. Here, the datastreams represent daily ecological momentary assessments (EMA) via a mobile App and temporal behavioral patterns collected from wristbands that are preassigned to the medical interns. As a medical internship--the first phase of professional medical training in the United States--is a stressful period in the career of physicians, the residents are faced with difficult decisions, long work hours and sleep deprivation. A timely identification of individuals with sleep- or activity-sensitive emotional states informs the policy maker right interventions, e.g,  the mobile App can send out sleep or activity message in the hope of promoting healthy outcomes. In this context, as new data batch arrives every day (such as hours of sleep, daily step counts and daily mood scores), our DTS framework quickly revises the statistical model and update the underlying parameter estimation, hence it enables an efficient detection of medical interns potentially at high stress level in a timely manner.

\subsection{Model setup and our contribution}

Driven by the aforementioned examples, we formulate the dynamic tracking and screening problem as follows. Suppose that we have $p$ datastreams for the units indexed by $j=1,\ldots, p$. Suppose the study begins at the time point $t_1$, and we are at the current time point $t_m$. At each point $t_i$, $i=1, \ldots, m$, we observe the response $y_{ij}$ and the covariates $\X_{ij}\in \mathbb{R}^d$.
We consider the model in the form of
\begin{align}\label{model1}
y_{ij}{=}\left\{\hspace{-0.1cm}
\begin{array}{ll} \X_{ij}^{\sT}\bbe(t_i)+\sigma(t_i)\varepsilon_{ij}, & {\mathrm {for\ }} \quad t_{i}\in(0,\tau_j],\\[0.1cm]
\X_{ij}^{\sT}\{ \bbe(t_i)+\bm\delta_j(t_i)\}+\sigma(t_i)\varepsilon_{ij}, & {\mathrm {for\ }} \quad t_{i}>\tau_j,
\end{array}\right.
\end{align}
for $j=1,\ldots, p$, $i=1,2,\ldots$, where $\varepsilon_{ij}$ is the random noise that satisfies $\E(\varepsilon_{ij}\mid {\bf X}_{ij})=0$ and $\var(\varepsilon_{ij}\mid {\bf X}_{ij})=1$ for theoretical treatments, $\sigma^2(\cdot)$ is the variance function, and $\tau_j$ is an unknown change-point in the $j$th stream. The coefficient $\bbe(\cdot)$, the drift $\bm\delta_j(\cdot)$, and the variance function $\sigma^2(\cdot)$ are assumed to be smooth functions. In model (\ref{model1}), different streams are assumed to share the same coefficient function $\bm\beta(\cdot)$, while the change-points $\tau_j$'s and the drift functions $\bm\delta_j(\cdot)$'s are allowed to vary among different streams.

To better understand our model in \eqref{model1}, take IHS for example, the response $y_{ij}$ is the mood score (self-reported through the mobile App), and the covariates $\X_{ij}$ include traits of individual which may affect the mood score, such as hours of sleep, step counts, or average resting heart rate. As the relationship between the mood score and the covariates is potentially affected by factors such as temperature or daylight hours that change with time, it is more suitable to treat the regression coefficients as dynamic functions of the time. In addition, the time-varying coefficient $\beta_r(\cdot)$ (i.e., the $r$th coordinate of $\bbe(\cdot)$) captures the mean change in the mood score if, for example, the individual sleeps one hour less while holding other predictors in the model constant, and therefore is shared across different streams. Lastly, since the medical interns work in high-stress environments, some of them may experience episodes of mood change after an initial period during which they adapt to daily routines. Such irregular patterns would often last for a period, which leads to (\ref{model1}).

Given Model \eqref{model1}, at the current time point $t_m$, our goal is two-folds: First, we want to provide accurate estimates of $\bbe(t_m)$ and $\sigma^2(t_m)$ without having to store the entire historical trajectory for each subject (Section \ref{Sec-2.1}). We refer to this parameter estimation step as the dynamic tracking step. Second, we aim to sequentially detect the occurrence of the changes as soon as possible {\it for each stream} (Section \ref{Sec-2.2}). We define, formally, $\mathcal{O}_{t_m} = \{ j:\ \|\bm\delta_j(t_m)\| \neq 0,\ j=1,\ldots,p \}\subset\{1,\ldots,p\}$ as the subset that contains the indices of the irregular datastreams at the current time $t_m$, where $\|\cdot\|$ represents the vector $L_2$ norm. We refer to this second change-point detection step as the dynamic screening step.

Although there have been some recent articles expressing concern about the online updating method for analysis of datastreams (see \cite{Schifano_2016, luo2020renewable} and the references therein), the issues of developing effective methodologies and theories for statistical modeling and inference of massive datastreams still remain. As most of the existing procedures and formulae were mainly developed based on the assumption that the observations come from the same model across time and sources.
The primary goal of this manuscript is to provide a dynamic statistical inference--dynamic tracking and screening (DTS)--procedure that fully explores the dynamical features of datastreams. We summarize our contribution from two perspectives.

From a statistical methodology standpoint, our dynamic tracking and screening (DTS) procedure efficiently adapts local dynamic structures in the streaming data environment and detect the irregular patterns as soon as they occur. Specifically, we demonstrate that incorporating exponentially weighted loss functions into our DTS procedure allows the estimates and the test statistics to be updated sequentially in a timely manner (Eqs. \eqref{betare} and \eqref{tup}). As a result, DTS can quickly revise the underlying statistical model as the new data arrive, without the need to store an ever-increasing data history (see Section~\ref{Sec-2.3} for a discussion on computational complexity).

From a theoretical perspective, our theoretical investigations show that the proposed estimators in DTS are uniformly consistent under some regularity conditions on between-and-within streams dependence even when the proportion of irregular streams does not vanish to zero as $p$ goes to infinity, and the optimal convergence rates of the proposed estimators are presented as separate results (Theorem \ref{thm2}). It is also worth noting that, different from the classical nonparametric estimation literature where the smoothing parameter is often chosen to minimize the global mean squared error, our DTS chooses the smoothing parameter adaptively so that the averaged prediction error is minimized (Theorem \ref{thm3}). With the help of efficient tracking, a new multiple testing procedure tailed to the streaming environment is developed for screening purposes, and we show that the false discovery rate (FDR) with the data-driven threshold can be controlled at the nominal level uniformly at all time points (Theorem \ref{thm4}).

\subsection{Connections to existing work}

Model (\ref{model1}) is built upon a varying coefficient (VC) model that incorporates potential structural changes raised by irregular patterns. The VC models have been extensively studied in the past two decades, especially in the field of longitudinal data analysis, and are known to be very powerful tools for analyzing the relationship between a response and a group of covariates due to its efficiency and flexibility; see \cite{Fan_2008} for a comprehensive review. The VC models are particularly useful to explore the dynamic pattern in our problem discussed above.  Nevertheless, our framework differs substantially from existing literature in various aspects. First, traditional methods usually assume that all of the longitudinal observations have the same model structure, which is not appropriate in the present problem. Second, as the data are collected sequentially and our aim is prospective inference rather than retrospective analysis, only partial information is available rather than the whole functional curves as in the standard VC models. We need to make decisions based on the available observations up to the current time, and with the implicit assumption that more recent observations are more useful for making decisions. Third, as the data collection process runs with high speed, a sequential procedure which is capable of updating parameters with minimal storage requirements is highly desirable. These differences play an important role in the setup of our approaches for parameter estimation and hypothesis testing.

There are some efforts to adapt various sequential change-detection methods to large-scale datastreams surveillance, such as %\cite{veeravalli2001decentralized},
\cite{tartakovsky2006novel}, %\cite{zou09},
\cite{Mei_2010}, \cite{xie2013sequential}, \cite{Zou_etal_2015},  \cite{chan2017optimal} and \cite{ren2020large}. Different from our goal in developing dynamic statistical inference, their settings are completely from ours because they aim to minimize the overall expectation delay while controlling the average run length under the null hypothesis that none of the datastreams experience changes.
%The dynamic screening appears to be a multi-stage or hierarchical multiple testing problem in the literature. \cite{benjamini2007false},  \cite{yekutieli2008hierarchical}, and \cite{goeman2010sequential} have all investigated the error control issue. However, existing methods
%are usually designed for different types of applications and cannot be easily tailored for the problem under consideration.
Recent works on sequential testing based on the sequential probability ratio
test (SPRT) rules such as \cite{bartroff2018multiple} and \cite{song2019sequential} are computationally intensive, making it infeasible for large-scale studies such as
those arising from IHS where millions of tests are conducted simultaneously at each time.

More closely related works are \cite{Marshall_etal_2004}, \cite{Grigg_etal_2009}, \cite{Spiegelhalter_etal_2012}, and \cite{Gandy_Lau_2013}, which considered various applications of Benjamini and Hochberg's FDR control procedure for statistical surveillance. Those methods usually assume that the coefficient function $\bbe(\cdot)$ is either a constant or fully known, which is not appropriate to assume in the present problem since we need to exploit the time-varying structure of (\ref{model1}) in a data-driven manner.
Recently, some authors considered dynamic testing systems to solve the curve monitoring problem, e.g., see \cite{qiu2014univariate} and \cite{qiu2015surveillance}, among others.
%In those methods, the regular longitudinal pattern is first estimated from the observed longitudinal data of a group of well-functioning subjects, and then a standard change detection scheme is applied to the standardized observations of a new subject for sequentially monitoring its longitudinal behavior. 
Those methods are simple and effective, however, they are designed under the assumption that the regular pattern is known in advance of detection and only focus on the irregular behavior of local streams, without taking the global false discoveries into account.
It should be also emphasized that to the authors' best knowledge, a common feature of most of the literature on sequential detection consists in the fact that either the datastreams are independent \citep{Gandy_Lau_2013,xie2013sequential} or the streaming observations are independent in the time domain \citep{ren2020large}. However, such assumption is often violated in practice, especially for large-scale datastreams with high frequency observations, which may in turn hamper their applicability to massive data applications. To this end, this article suggests a systematic DTS procedure by making connections to the VC models and some sequential detection problems. We address two key challenges in a unified framework: constructing efficient estimators and multiple testing procedures under model (\ref{model1}), and investigating theoretical properties under lenient requirements on the datastreams.

\subsection{Organization}
 Our paper is structured as follows. In Section~\ref{Sec-2}, we propose the dynamic tracking and screening procedure, followed by investigating its theoretical properties in Section~\ref{Sec-3}. In Section~\ref{Sec-4}, we conduct simulation experiments to show the finite-sample performance of the proposed method in comparison with some others, and then apply our approach on the Intern Health Study for further illustration in Section~\ref{Sec_IHS}. Section~\ref{Sec-5} offers a summarizing discussion. The proofs and some technical details are delineated in the Supplementary Material.

\section{Methodology}\label{Sec-2}
Before proceeding, to avoid confusion, we would like to note that we use $t_m$ to represent the current time point, and use $t_i$ to represent a given time point prior to current time point $t_m$.
At the current time point, we have access to the observations up to time point $t_m$, i.e., $\{(y_{1j},\X_{1j}),\ldots,(y_{mj}, \X_{mj})\}$.
In addition, because our over-aching goal is to dynamically identify individuals with irregular behaviors and to estimate $\bbe(t_m)$ captures the shared time-varying coefficient across all individuals,  we work under the setting that individuals with the irregular pattern $\bm\delta_j(t_m)$ are minority in the study cohort. In other words, the cardinality of $\mathcal{O}_{t_m}$ is small compared to $p$.

\subsection{Dynamic Tracking}\label{Sec-2.1}

We start with describing our approach for dynamic tracking on estimating $\bbe(t_m)$.
Because different datastreams have different time-varying coefficients (some have $\bbe(t_m)$ and some have $\bbe(t_m) + \bm\delta_j(t_m)$, unless the irregular set $\mathcal{O}_{t_m}$ is known as a prior, we cannot naively combine all observed data $\{(y_{1j},\X_{1j}),\ldots,(y_{mj}, \X_{mj})\}$ together to construct an accurate estimator of $\bbe(t_m)$. Instead, at current time $t_m$, our dynamic tracking strategy for estimating $\bbe(t_m)$ is to first estimate each individual stream coefficients and then perform a robust quantile-based approach to combine those estimates after screening out the irregular ones.

For each datastream $j$, at the current time $t_m$, based on the observed data $\{(y_{1j},\X_{1j}),\ldots,(y_{mj}, \X_{mj})\}$ up to the time point $t_m$, we consider the following loss function to better incorporate local dynamics:
\begin{align}\label{lf}
Q_{mj,\lambda}({\bf b}) = \sum\limits_{i=1}^m\left(y_{ij}-\X_{ij}^{\sT}{\bf b}\right)^2\lambda^{t_m-t_i} \triangleq \sum\limits_{i=1}^m\left(y_{ij}-\X_{ij}^{\sT}{\bf b}\right)^2w_i(t_m),\quad j=1,\ldots,p,
\end{align}
where $\lambda\in(0,1)$ is a smoothing parameter (see Section \ref{Sec-2.3} for its choice in practice), and $w_i(t_m)$ assigns different weights to different individuals.

The exponential weighting function $w_i(t_m)$ invests at least two merits into our framework. First, it respects the local dynamic nature of our streaming data environment. Because the weighting function assigns smaller weights to individuals farther away from the current time point $t_m$, and all observations up to $t_m$ are incorporated in $Q_{mj,\lambda}({\bf b})$ to improve statistical estimation efficiency. We note that $Q_{mj,\lambda}({\bf b})$ combines the ideas of local smoothing and exponential weighting schemes used in the exponentially weighted moving average (EWMA) procedures through the term $\lambda^{t_m-t_i}$, which can be regarded as weights defined by a special kernel function \citep{Runger_1996}. Second, the exponential weighting function admits recursive expressions in both tracking and screening steps, whereas some commonly used kernels, such as the Epanechnikov kernel $K(u)=0.75(1-u^2)_{+}$, may not result in recursive formulae. In other words, though this weighting scheme is just one choice among a broad class of weighting functions, it serves the purpose of our DTS procedure well. In addition, as opposed to the traditional VC model, making one-sided kernel functions is necessary in our problem, since only the observations on one side of $t_m$ are available \citep{Wu:1993ec}.

Benefited from the exponential weighting function $w_i(t_m)$, we can quickly obtain an estimate of $\bm\beta(t_m)$ that not only minimizes the lost function $Q_{mj,\lambda}({\bf b})$ but also can be updated efficiently based on the previous time point estimate $\widehat{\bbe}_{j,\lambda}(t_{m-1})$:
\begin{align}\label{betare}
\widehat{\bbe}_{j,\lambda}(t_m)&=\underset{b\in\mathbb{R}^d}{\arg\min} \ Q_{mj,\lambda}({\bf b}) = {\bf A}_{mj}^{-1}\left\{w_{m-1}(t_m){\bf A}_{m-1,j}\widehat{\bbe}_{j,\lambda}(t_{m-1})+\X_{mj}y_{mj}\right\},\\
{\bf A}_{mj}&=w_{m-1}(t_m){\bf A}_{m-1,j}+\X_{mj}\X_{mj}^{\sT}\nonumber.
\end{align}
Then, by defining $e_{ij}=y_{ij}-\X_{ij}^{\sT}\widehat{\bbe}_{j,\lambda}(t_i)$, the variance function $\sigma^2(\cdot)$ at $t_m$ can be estimated using the observations on the $j$th stream by
\begin{align}\label{siup}
\widehat{\sigma}_{j,\lambda}^2(t_m) =\varphi_m^{-1}\sum_{i=1}^m w_i(t_m)e_{ij}^2,
\end{align}
which again can be recursively updated:
\begin{align}
\widehat{\sigma}_{j,\lambda}^2(t_m)=\varphi_m^{-1}\left\{w_{m-1}(t_m)\varphi_{m-1}\widehat{\sigma}_{j,\lambda}^2(t_{m-1})+e_{mj}^2\right\},
\end{align}
with $\varphi_m=\sum_{i=1}^mw_i(t_m)$. In Section~\ref{Sec-3}, we prove in Theorem \ref{thm1} that $\widehat{\bbe}_{j,\lambda}(t_m)$ and $\widehat{\sigma}_{j,\lambda}^2(t_m)$ are appealing estimates given a properly chosen $\lambda$.

In the second step, because the majority of $\widehat{\bbe}_{j,\lambda}(t_m)$'s are correctly centered at $\bbe(\cdot)$ and the others substantively deviate from $\bbe(\cdot)$ pushed by the irregular pattern $\bm\delta_j(t_m)$. This motivates us to adopt a robust quantile-based method to better estimate $\bbe(\cdot)$ and $\sigma^2(\cdot)$. 
We estimate $\bm \beta(t_m)$ component-wise. For the $r$th component $\beta_r(t_m)$,
the signal set $\mathcal{O}_{r, t_m}=\left\{ j: \ \delta_{jr}(t_m) \neq 0 \right\} \subset \{1,\ldots, p\}$ can be divided into two subsets
\begin{align*}
\mathcal{O}^{+}_{r, t_m} \cup \mathcal{O}^{-}_{r, t_m}
:=
\left\{ j: \ \delta_{jr}(t_m) > 0 \right\} \cup  \left\{ j: \ \delta_{jr}(t_m) < 0 \right\}  ,
\end{align*}
where $ \mathcal{O}^{+}_{r, t_m} $ contains the subjects with positive biases and  $\mathcal{O}^{-}_{r, t_m}$ includes the rest. By definition, $\mathcal{O}_{t_m}=\cup_{r=1}^{d} \mathcal{O}_{r, t_m}$.
Accordingly, we define
\begin{align*}
&\pi_{r, t_m}^{+} = \frac{ \Card(\mathcal{O}^{+}_{r,t_m}) }{ \Card(\mathcal{O}_{r,t_m}) },\quad
\pi_{r, t_m}^{-} = \frac{ \Card(\mathcal{O}^{-}_{r,t_m}) }{ \Card(\mathcal{O}_{r,t_m})  },\quad
\pi_{r, t_m} = \frac{1}{2} - \frac{ \pi_{r, t_m}^{+}-\pi_{r, t_m}^{-} }{2},
\end{align*}
with $\Card(\cdot)$ being the cardinality of any set.
Let $\hat{\beta}_{jr, \lambda}(t)$ be the $r$th component of $\hat{\bs{\beta}}_{j, \lambda}(t)$, for $r=1, \ldots, d$.
We estimate $\beta_r(t_m)$ through the $\pi_{r, t_m} $-th quantile of  $ \hat{\beta}_{1r,\lambda}(t_m), \cdots,$ $ \hat{\beta}_{pr,\lambda}(t_m)$, i.e.,
\begin{align}\label{b0e}
\tilde{\beta}_{r,\lambda}(t_m) = \inf \left\{ \beta_r : \ \pi_{r, t_m} \leq  \hat{F}_p(\beta_r,t_m)  \right\},
\end{align}
where $\hat{F}_p(\beta_r,t_m)=p^{-1}\sum_{j=1}^{p}\ID( \hat{\beta}_{jr,\lambda}(t_m)\leq \beta_r )$ and $\bI(\cdot)$ is an indicator function. We note that $ \pi_{ r, t_m }$ is unknown but it can be estimated consistently, and the estimation details shall be provided in Section~\ref{Sec-2.3}.

Constructing an estimate of $\sigma^2(t_m)$ using the information of all datastreams is much simpler because the consistency of $\widehat{\sigma}_{j, \lambda}^2(t_m)$ in (\ref{P.2}) can always be obtained, regardless of whether $j$ is an outlying stream. Naturally, we estimate $\sigma^2(t_m)$ by the pooled average from all datastreams, which is $\tilde{\sigma}^2_{\lambda}(t_m)=p^{-1}\sum_{j=1}^p\widehat{\sigma}_{j, \lambda}^2(t_m)$.

Like many other smoothing-based procedures, setting the value of the smoothing parameter $\lambda$ is a critical and non-trivial task. A larger $\lambda$ may gain on the variance side, but loses on the bias side. The optimal choice of $\lambda$ often depends on the ``smoothness" of $\bbe(t_m)$, a quantity that is only estimable under strict, often unrealistic, assumptions. Cross-validation has been frequently adopted in selecting the bandwidth in the VC models or longitudinal data analysis literature \citep{Hoover_1998}. The selection of an optimal $\lambda$ is particularly challenging in the present problem as we need to determine it dynamically for the current time, say $\lambda(t_m)$. This is because the population coefficient function typically varies with time and it does not make sense to assume that $\bbe(t_m)$ and $\bbe(t_{m'})$ have the same degrees of smoothness if $|t_m-t_{m'}|$ is large. This implies that the optimality of estimation across all the time points simultaneously cannot be achieved with a single choice of $\lambda$ \citep{zhang00}.

\textit{Choice of the smoothing parameter.} Note that at the current time point $t_m$, we are only concerned about the estimation of $\bbe(t_m)$ rather than the entire regression coefficients; that is, in our dynamic tracking procedure we do not update the estimation of $\bbe(t)$ for $t<t_m$ after receiving the streaming observation at $t_m$. As a result,
it is reasonable to define the averaged predictive squared error (APSE) of $\tilde{\bbe}_{\lambda}(t_m)$ by
\[
\APSE_{\lambda}(t_m) = \Card(\mathcal{I}_{t_m})^{-1}\sum_{j\in \mathcal{I}_{t_m}}\{y_{mj}^{*}-\bX_{mj}^\top\tilde{\bs{\beta}}_{\lambda}(t_m)\}^2,
\]
where $\mathcal{I}_{t_m}\subset\{1,\ldots, p\}$ is a subset that contains most of the noise streams, $y_{mj}^{*}$ is a new observation at $(\X_{mj},t_{m})$, say $y_{mj}^{*}=\X_{mj}^{\top}\bbe(t_m)+\varepsilon_{mj}^{*}$, where $\varepsilon_{mj}^{*}$ is a new realization of $\varepsilon_{ij}$. Note that $\mathcal{I}_{t_m}$ is allowed to contaminate with some outlying streams provided that its size is small relative to $p$. We shall provide a heuristic algorithm to select $\mathcal{I}_{t_m}$ in Section~\ref{Sec-2.3}.
By the smoothness assumption of $\bbe(\cdot)$, we propose a one-step APSE criterion to choose $\lambda$ dynamically from
\begin{align}\label{lhe}
\hat{\lambda}(t_m) =\mathop{\arg\inf}_{\lambda} \hat{\APSE}_{\lambda}(t_m),
\end{align}
where $\hat{\APSE}_{\lambda}(t_m) = \Card(\mathcal{I}_{t_m})^{-1}\sum_{j\in \mathcal{I}_{t_m}}\{y_{mj}
-\bX_{mj}^\top\tilde{\bs{\beta}}_{\lambda}(t_{m-1}) \}^2$.

\subsection{Dynamic screening}\label{Sec-2.2}

Now, let us turn to construct an effective screening procedure. By defining the normalizing transformation of the response $y_{ij}$ as  $z_{ij}=\{y_{ij}-\X_{ij}^{\sT}{\bbe}(t_i)\}/{\sigma}(t_i)$, we have under model \eqref{model1}
\begin{align}\label{model31}
z_{ij}{=}\left\{\hspace{-0.1cm}
\begin{array}{ll} \varepsilon_{ij},  & {\mathrm {for\ }} \quad t_i\in(0,\tau_j],\\ [0.1cm]
\gamma_j(t_i)+\varepsilon_{ij}, & {\mathrm {for\ }} \quad
t_i>\tau_j,
\end{array}\right.
\end{align}
where $\gamma_j(t_i)=\X_{ij}^\top \bm\delta_j(t_i)/\sigma(t_i)$, and $\bm\delta_j(\cdot)$ is a smooth function defined on $(\tau_j,\infty)$, for $j=1,\cdots, p$, $i=1,2,\cdots$. In this context, the goal of change points detection is to sequentially check if $z_{ij}$ has zero mean at each time point. More formally, at current time $t=t_m$, we test the null hypotheses
\begin{align}
H_{mj}^0: \ \E( z_{mj} )=0,\quad j=1,\ldots,p.
\end{align}
As $\bbe(t_i)$ and $\sigma^2(t_i)$ can be well estimated by $\tilde{\bbe}_{{\lambda}}(t_i)$ and $\tilde{\sigma}_{\lambda}^2(t_i)$, respectively, $\z_{ij}=\{y_{ij}-\X_{ij}^{\sT}\tilde{\bbe}_{\lambda}(t_i)\}/\tilde{\sigma}_{\lambda}(t_i)$ yields a natural estimate of $z_{ij}$. Similar to the spirit of \eqref{betare}, we estimate $\gamma_j(t_m)$ through
\begin{align}\label{lceg}
\widehat{\gamma}_{j,\lambda}(t_m)& := \varphi_m^{-1}\sum_{i=1}^mw_i(t_m)\z_{ij},
\end{align}
which serves as a proper quantity at current time point $t_m$ for checking $H_{mj}^0$. The dynamic screening procedure rejects the null hypothesis whenever the streaming pattern of $j$th datastream deviates from that of the majority, i.e.,  if $|\widehat{\gamma}_{j,\lambda}(t_m)|$ exceeds a pre-specified threshold  at $t_m$.

Again, benefiting from the exponential weight function $w_i(t_k)$, we have the following recursive form to quickly update $\hat{\gamma}_{j,\lambda}(t_m)$ as the new data arrive
\begin{align}\label{tup}
\widehat{\gamma}_{j,\lambda}(t_m)&=\varphi_{m}^{-1}\left\{w_{m-1}(t_m)\varphi_{m-1}\widehat{\gamma}_{j,\lambda}(t_{m-1})+\z_{mj}\right\},
\end{align}
where $\varphi_m=\sum_{i=1}^mw_i(t_m)$. Under the null hypotheses and certain regularity conditions, we prove that $\widehat{\gamma}_{j,\lambda}(t_m)$ is asymptotically normal. %{\color{red} %(Supplementary Material Lemma \ref{UC})}.
Nevertheless, because reliable estimates of $\widehat{\gamma}_{j,\lambda}(t_m)$'s long-run variance are extremely challenging to obtain in the presence of those irregular patterns,  multiple testing procedures that directly adjust for p-values are not suitable to achieve our goal.

%Thanks to the essence of our dynamic screening procedure is to maintain high identification accuracy without excessive false positives.
We consider an alternative way to apply the well known \cite{Benjamini_Hochberg_1995}'s FDR procedure (BH) with statistical guarantees (Theorem \ref{thm4}).
Concretely, we reject $H_{mj}^0$ if $|\widehat{\gamma}_{j,\lambda}(t_m)|>L$, where $L$ is a data-driven threshold \citep{Storey_etal_2004} given by
\begin{equation}\label{drl}
L = \inf\left\{u: \frac{\#\{j: |\widehat{\gamma}_{j,\lambda}(t_{*})| \ge u  \}  }{\#\{j: |\widehat{\gamma}_{j,\lambda}(t_m)| \ge u  \}\vee 1 }\le \alpha    \right\}
\end{equation}
for a desired FDR level $\alpha$ and some small $t_*>0$. The screening set is denoted by $\widehat{\mathcal{O}}_{t_m}$.  In (\ref{drl}), we use a ``warm-up" sample to construct a series of null test statistics, denoted as $\widehat{\gamma}_{j,\lambda}(t_*), j=1, \ldots, p$. Such a warm-up sample is generally available since a conventional assumption in the practice of change-detection is that the changes would be unlikely to occur at the beginning of monitoring. {\color{black}In our motivating example, this is also a reasonable assumption in the context of regular medical residence training where work and life routines are being established as baseline from which deviations may be detected.} Intuitively, if most of $|\widehat{\gamma}_{j,\lambda}(t_*)|$'s are from null states, $\#\{j: |\widehat{\gamma}_{j,\lambda}(t_*)| \ge u \}$ would be a reasonable approximation to $\#\{j: |\widehat{\gamma}_{j,\lambda}(t_m)| \ge u,H_{mj}^0\}$. The advantage of this empirical formula is that we do not need to estimate other nuisance parameters and the temporal correlation structures are allowed to be different across datastreams. In Section~\ref{Sec-3}, we shall prove that, under mild conditions, the false discovery proportion (FDP), defined as
\begin{align}\label{fdpl}
\FDP(t_m):= \frac{\#\{j: \widehat{\gamma}_{j,\lambda}(t_m) \ge L, j \notin \mathcal{O}_{t_m} \}  }
{\#\{j: \widehat{\gamma}_{j,\lambda}(t_m) \ge L  \}\vee 1 }
\end{align}
with the threshold $L$ is asymptotically controlled at the level $\alpha$, and such control is valid uniformly at $t_m$.

\subsection{Practical implementations}\label{Sec-2.3}

In this subsection, we summarize our DTS procedure and discuss some issues for its implementation. Our proposed DTS procedure is summarized as follows.
\vspace{0.2cm}

{\it Dynamic Tracking and Screening (DTS) Procedure}
\begin{itemize}
	\item[1.] (Initiation) Set $\Lambda=\{\lambda_k,k=1,\ldots, q\}$ and an FDR level $\alpha$.
	
	\item[2.] (Choice of the smoothing parameter) Given the observations \\ $\{(y_{mj},\X_{mj})\}_{j=1}^p$ at the time point $t_m$, find the optimal $\lambda$ by
	$\widehat{\lambda}(t_m)=\arg\inf_{\lambda\in\Lambda}\widehat{\mbox{APSE}}_{\lambda}(t_m)$.
	
	\item[3.] (Dynamic tracking) Update the estimators $\widehat{\bbe}_{j,\lambda}(t_m)$ and $\widehat{\sigma}_{j,\lambda}^2(t_m)$ by (\ref{betare}) and (\ref{siup}) for each $\lambda\in\Lambda$. Obtain $\tilde{\bm\beta}_{\widehat{\lambda}(t_m)}(t_m)$ by (\ref{b0e}) and $\tilde{\sigma}^2_{\hat{\lambda}(t_m)}(t_m)$ (the implementation details are provided in the following.
	
	\item[4.] (Dynamic screening) Compute $\z_{mj}$ %and $s(t_m,t_i)$ for $\{i:t_m-t_i\leq \omega\}$. Obtain the decorrelted residual $\tilde{\varepsilon}_{mj}$ by the transformation given in Section 2.3.1.
	and calculate the test statistics $\widehat{\gamma}_{j, \lambda(t_m) }(t_m)$ for $j=1,\ldots,p$ using (\ref{tup}).
	Search for the threshold $L$ by (\ref{drl}); Display the discoveries with the level $\alpha$, i.e., $\widehat{\mathcal{O}}_{t_m}$.
	
\end{itemize}
In step 2, since $\widehat{\lambda}(t_m)$ can only be approximately identified within a  compact set of the parameter space and there might exist more than one local minimum,
in practice we recommend to find $\widehat{\lambda}(t_m)$ from a pre-specified set with some admissible values, say $\Lambda=\{\lambda_k,k=1,\ldots, q\}$.
%We note that the recursive form (\ref{betare}) is not applicable when $\lambda$ is time-varying
%Then, at each point $t$, we can update the estimates with the recursive forms.

\vspace{0.3em}
\noindent\textit{Discussion on computation complexity}
In our DTS procedure, we  need to recursively store ${\bf A}_{mj}$, $\widehat{\bm\beta}_{j,\lambda}(t_m)$, $\hat{\sigma}_{j,\lambda}(t_m)$ and $\hat{\gamma}_{j, \lambda}(t_m)$. With the help of the recursive formulae, the computational complexity at each time point is linear in $q$ and $p$ and does not depend on $m$. Although updating $\hat{\bbe}_{j,\lambda}(t_m)$ from (\ref{betare}) requires a matrix inverse calculation, we can alternatively apply the Plackett updating formula in \cite{Harville_1997} to obtain a fast update of this inversion, as the perturbation $\X_{mj}\X_{mj}^{\sT}$ in ${\bf A}_{mj}$ is a rank-one matrix. Thus, the storage space required for our procedure is of the order $O(pqd^2)$. We also note that the parameter estimation for $p$ datastreams can be independently carried out, which suggests the computation burden may be further reduced with the help of parallel and distributed computing platforms. The R codes that implement the proposed scheme are available upon request.
% in the Supplementary Material.

\vspace{0.3em}
\noindent\textit{Determination of $\pi_{r, t_m}^{+}-\pi_{r, t_m}^{-}$ and $\mathcal{I}_{t_m}$}
To find the quantile-based estimate of the regression coefficient at $t_m$, in this section, we provide an estimate of the difference in proportions of positive and negative drifts (i.e., $\pi_{r, t_m}^{+}-\pi_{r, t_m}^{-}$).
%In practice, as the effect of $\delta_j(t_m)$ on estimating $\beta_{jr}(t_m)$ can be different for different $r$, we need to estimate $\pi_{t_m}^{(1)}-\pi_{t_m}^{(2)}$ (thus $\tau_{\pi_{t_m}}$) separately for each direction of $\bbe_{j}(t_m)$.
For a given component $r$ and a properly chosen $\lambda$, for $j \notin \mathcal{O}_{r, t_m}$, since $\hat{\beta}_{jr, \lambda}(t_m)$ is a weighted average that approximately centers around $\beta_{r}(t_m)$, we might expect that $\{\hat{\beta}_{jr, \lambda}(t_m)-\beta_{r}(t_m), \ j \notin \mathcal{O}_{r, t_m} \}$ tend to reside symmetrically on both side of 0, for $r=1, \ldots, d$. Then the set $\{j:\ \hat{\beta}_{jr, \lambda}(t_m)-{\beta}_{r, \lambda}(t_{m-1})>0, j=1,\cdots,p\}$ approximately contains the half of the regular subjects and the subjects with positive biases. As $\tilde{\beta}_{r, \lambda}(t_{m-1})$ is often a good estimate of  $\beta_{r}(t_m)$, naturally, we may use  $\#\{\hat{\beta}_{jr, \lambda}(t_m)-\tilde{\beta}_{r, \lambda}(t_{m-1})>0\}/p$ as an approximation of $\pi_{r, t_m}^{(0)}/2 + \pi_{r, t_m}^{+}$, and $\#\{\hat{\beta}_{jr, \lambda}(t_m)-\tilde{\beta}_{r, \lambda}(t_{m-1})<0\}/p$ as an approximation of $\pi_{r, t_m}^{(0)}/2 + \pi_{r, t_m}^{-}$, with $\pi_{r, t_m}^{(0)} = 1-\pi_{r, t_m}^{+}-\pi_{r, t_m}^{-}$. Therefore, our DTS procedure estimates $\pi_{r, t_m}^{+}-\pi_{r, t_m}^{-}$ for the $r$th direction through
$$
\#\{\hat{\beta}_{jr, \lambda}(t_m)>\tilde{\beta}_{r, \lambda}(t_{m-1})\}/p-\#\{\hat{\beta}_{jr, \lambda}(t_m)<\tilde{\beta}_{r, \lambda}(t_{m-1})\}/p.
$$
We provide the theoretical property of the above estimator in Proposition \ref{prop1}.

Since $\mathcal{I}_{t_m}$ is required to contain as few outlying streams as possible, and should be stochastically independent from the current observations updated at time $t_m$ from the proof of Theorem \ref{thm3}, motivated by the least trimmed squares in classical outlier detection \citep{rousseeuw1984least}, we suggest to use
$\mathcal{I}_{t_m}=\{j: |\hat{\gamma}_{j,\widehat{\lambda}(t_{m-1})}(t_{m-1})| \leq \gamma_{ [p/2] }\}$, where $\gamma_{[p/2]}$ is the $[p/2]$th smallest value in $\{ |\hat{\gamma}_{j,\widehat{\lambda}(t_{m-1})}(t_{m-1})|: j=1,\ldots,p\}$.
In this case, we implicitly assume that $\mathcal{O}_{t_{m-1}}$ and $\mathcal{O}_{t_m}$ do not differ too much, which is usually reasonable in practice. As $\{\hat{\gamma}_{j,\widehat{\lambda}(t_{m-1})}(t_{m-1})\}$ is a good measure to quantify deviations of the $j$th stream from the regular pattern, we may expect that $\mathcal{I}_{t_m}$ is a clean set without many irregular datastreams.
%The matrix $\widehat{\bm\Sigma}^{-1}({t_m})$ may not be semi-positive definite, whereas some simple adjustment procedure can
%effectively regularize it to be a well-defined covariance matrix; see for example Li (2011).

\vspace{0.3em}

\noindent\textit{Implementation in the presence of substructures} The model (\ref{model1}) assumes that all the datastreams share a common varying coefficient structure before the change occurs. This setup is commonly adopted in the literature of longitudinal/functional data analysis in which the regression function is supposed to be the same across the observed individuals; see %\cite{henderson2008nonparametric},
\cite{zhu2012multivariate} and \cite{yao2013new} among many others. This assumption, however, could be violated in some applications, especially
when the number of datastreams is extremely large.  In a wide range of cases, it may be more plausible  to suppose that there are groups of individuals who share the same regression function (or at least have very similar regression curves). As a modelling strategy, we may thus assume that the observed streams can be grouped into a number of classes whose members all share the same regression function. If the group information can be known as {\it a priori} given some auxiliary covariates (such as the age, professionals and some others in the IHS example), the DTS procedure is directly applicable for each group individually. Otherwise, we may employ some structure identification or classification methods developed in recent literatures on the warming-up dataset. Please refer to \cite{james2003clustering}%, \cite{chiou2007functional}
and \cite{ke2016structure} for model-based clustering approaches, and \cite{abraham2003unsupervised} and \cite{vogt2017classification} for some model-free methods.

Testing whether model (\ref{model1}) holds for all the streams or some given groups can be viewed as  the comparison of a large number of regression curves. This has been the object of much work, see for instances, \cite{Neumeyer:2003jo}, %\cite{PardoFernandez:2007ue}
\cite{wang2017comparison}, and \cite{GonzalezManteiga:2013fs} for a survey.

\section{Theoretical investigations}\label{Sec-3}
In this section, we derive the asymptotic properties of the proposed estimators. We first discuss the assumptions that are needed for the analysis and then summarize the main theorems. The proofs are provided in the Supplemental Material.

As we are considering the problem of dynamic tracking in an unending sequence, it is necessary to modify the traditional asymptotic regime used in nonparametric regression to better fit the current situation. Formally, we assume that $t_i$ is deterministic on a bounded support, which without loss of generality equals $[0, 1]$. Let $\eta>0$ be a basic time unit in a given application, which is the largest time unit that all (unequally spaced) observation times are its integer multiples. Define $n_i=t_i/\eta$, for $i=0, 1,\cdots$, where $n_0=t_0=0$. Then, $t_i=n_i\eta$, for all $i$, and $n_i$ is the $i$th observation time in the basic time unit. To mimic the unending sequence of the time points, we assume that there exists a huge $N \rightarrow \infty$ that denotes the ending of indefinitely long and time-evolving sequence such that $N\eta$ is bounded, say $N\eta=1$. In this regime, as $N$ increases, the $\eta$ goes to zero, and the current time point $t_m$ may be much small than the ending point that $0< t_m< 1$.
%Let $I_m=[t_{*}, t_m]$, for some small $t_{*}>0$.
\subsection{Assumptions}

We make the following assumptions to establish the theoretical foundation of our DTS procedure.
\begin{assumption} \label{A1}
The time series processes $\mathcal{D}_j=\{\bX_{ij}, \varepsilon_{ij}, i=1,\ldots, m\}$
are strictly stationary and strongly $\rho$-mixing for each $j$. Let $\rho_j(l)$ for $l=1,2, \ldots$ be the mixing coefficients corresponding to the $j$th time series $\mathcal{D}_j$. It holds that $\rho_j(l) \le \rho(l)$ for all $1\le j \le p$, where the coefficients $\rho(l)$ satisfy that $\sum_{l}\rho(l)<\infty$. Moreover, assume that the eigenvalues of $\bs{\Gamma}_j:=\mathbb{E}(\bX_{ij}\bX_{ij}^\top)$ are uniformly bounded by zero and infinity.
\end{assumption}

\begin{assumption} \label{A2}
Suppose $\{\mathcal{D}_j, j=1, \cdots, p\}$ satisfy the block dependence structure in the sense that
there exists a partition of the data streams $\{\mathcal{D}_{j, k}, j=1, \cdots, J, k=1, \cdots, n_j\}$
such that $\mathcal{D}_{j_1, k_1}$ are independent with $\mathcal{D}_{j_2, k_2}$ for $j_1 \ne j_2$, and the maximal block size in each partition is of the order $O(N/J)$. In addition, we assume that $J=p^{\xi}$, for some $\xi > 0$.

\end{assumption}

\begin{assumption} \label{A3}
There is a real number $\theta > 20/3$ such that $\sup_{i, j}\mathbb{E}\{ |\varepsilon_{ij}|^{\theta}\} <\infty$.
%, where $C$ is a fixed positive constant.
	
\end{assumption}
%\begin{assumption} \label{A4}
%For $i, i_1, i_2=1,\ldots,m$, $j=1,\ldots,p$, $\forall t_i, t_{i_1}, t_{i_2} \in I_m$, let
%	\begin{align*}
%  \bs{\Gamma}(t_i) = \mathbb{E}\left[\bX_{ij}\bX_{ij}^\top\right],\
%  \bs{\Gamma}(t_{i_1}, t_{i_2}) = \mathbb{E}\left[\bX_{i_1 j}\bX_{i_2j}^\top \right].
%	\end{align*}
%Assume for $l,r=1,\ldots,k$, $\bs\Gamma_{lr}(t_i)$ and $\bs{\Gamma}_{lr}(t_{i_1}, t_{i_2})$ are continuous and bounded on $I_m$.
%\end{assumption}

\begin{assumption} \label{A4}
	For $t\in [t_*, t_m]$, the varying coefficients $\beta_r(t)$ and $\delta_{jr}(t)$ for $r=1, \ldots, k$ and the variance function $\sigma^2(t)$ are twice differentiable with bounded derivative.
\end{assumption}

\begin{assumption} \label{A5}
	The number of data streams $p$ diverges to infinity that $p=O(1/\eta)$. The tuning parameter $\lambda$ has the property that $-1/\log \lambda \rightarrow 0$, $(-1/\log\lambda)/\{\eta\log(1/\eta) \} \rightarrow \infty$ and $C_1(1/\eta)^{-2/5}\le -1/\log \lambda \le C_2(1/\eta)^{-\delta}$, for some
	small $\delta>1/3$ and positive constants $C_1$ and $C_2$.
\end{assumption}

%\begin{assumption}\label{A6}
%Assume $|\delta_{jr}(t)|/\sqrt{-\log(p)\eta \log(\lambda)} \rightarrow \infty$, for $j \in \mathcal{O}_{r, t_m}$ and $t>\tau_j$.
%\end{assumption}

%\begin{assumption} \label{A7}
%	For $t\in I_m$, $j\not\in \mathcal{O}_t$, let $U_{jr}(t) = \sqrt{2(1/\eta)(-1/\log \lambda)}\{\hat{\beta}_{jr, \lambda}(t)- \beta_r(t)\}$, and $F_{U_{jr}(t)}(\cdot)$ be the cumulative distribution function of $U_{jr}(t) $. Assume that $F_{U_{jr}(t)}(\cdot)$ is symmetric with respect to zero and
%\begin{align}\label{eq:assump7}
%& |F^{\prime}_{U_{jr}(t)}(x) | > \nu >0,\ \forall |x|\leq K/\sqrt{p_0},
%\end{align}
%for some $\nu>0$ and $K>0$.
%\end{assumption}

The condition on the mixing rate $\rho(l)$ in Assumption~\ref{A1} is not
stringent and it can be satisfied if $\rho(l)$ decays to zeros by sufficiently high polynomial rates. Assumption~\ref{A2} implies that any data stream can be correlated with at most other $O(p^{1-\xi})$ data streams. The moment conditions in Assumptions~\ref{A3} are used to derive the uniform consistency of the regression coefficients; see \cite{Hansen08} for similar assumptions.
Assumption~\ref{A4} is the standard smoothness condition commonly used in the literature of VC model. Assumption~\ref{A5} imposes restriction on the relative growth of $p$ and the time unit $\eta$, and the smoothing parameter $\lambda$ is chosen at a rate faster than the optimal rate in the standard nonparametric regression problems so that the bias term will be negligible. Moreover, the lower bound rate $-2/5$ is closely related to the moment condition $\theta>20/3$ in Assumption \ref{A3}; see \cite{vogt2017classification} for similar discussions.
%Assumption~\ref{A6} guarantees that the signals of the alternative streams can be identified so that the quantile-based estimator will be consistent.
%Assumption \ref{A7} contains sufficient conditions for the point-wise consistency of $\tilde{\bm \beta}_{\lambda}(t)$. For $j\notin \mathcal{O}_t$, as $U_{jr}(t)$ is a weighted average that approximately centers around zero,  the asymptotic symmetry property of $U_{jr}(t)$ can be studied at the expense of more complicated proofs. In the present paper, strict symmetry is assumed in Assumption \ref{A7} to simplify theoretical derivations. The condition in \eqref{eq:assump7} is a standard assumption in quantile-based estimation procedures, which says the density of $U_{jr}(t)$ is bounded away from zero in a small neighborhood of its population median.

\subsection{Main results}

We first discuss the asymptotic properties of $\widehat{\bbe}_{j,\lambda}(t)$ and $\widehat{\sigma}_{j,\lambda}^2(t)$, for $t\in [t_*, t_m]$, where $[0, t_{*}]$ serves as a warm-up period. %We introduce a few additional notations for convenience. Let $c_{j,B}(t) = \mathbb{E}\left\{ h^{-1}\exp\{-(t-t_i)/{h}\}\ID(t_i \le t) \bm\Gamma(t_i) \bm\delta_j(t_i) \right\}$ with $h=-1/\log \lambda$, then $c_{j,B}(t) = 0$, for $j\notin \mathcal{O}_t$, and $\bm\Gamma(t)f(t) c_{j,B}(t) $ represents the bias of $\hat{\beta}_{jr,\lambda}(t)$ for $j\in\mathcal{O}_t$.
We assume that $h= -1/\log(\lambda) \to 0$ (say, $\lambda\to 0$) as $\eta\to 0$ in the subsequent analysis. Throughout this paper, we assume $\Card(\mathcal{O}_{t_m}) \leq c p$ for some $c<1$, which includes the sparse setting $\Card(\mathcal{O}_{t_m}) =o(p)$. Theorem \ref{thm1} establishes the uniform consistency of $\widehat{\bbe}_{j,\lambda}(t)$ and $\widehat{\sigma}_{j,\lambda}^2(t)$  for all $t\in [t_{*}, t_m]$.

\begin{theorem}\label{thm1}
	Under Assumptions \ref{A1} and \ref{A3}-\ref{A5}, $\widehat{\bbe}_{j,\lambda}(t)$ and $\widehat{\sigma}_{j,\lambda}^2(t)$ satisfy
	\begin{align}
	&\sup_{t \in [t_{*}, t_m] } \|\hat{\bm\beta}_{j,\lambda}(t)-\{\bm\beta(t)+\bm\delta_{j}(t)\} \|
	=O_p(h+ \sqrt{\log (1/\eta)\eta/h }), \label{P.1}\\
	& \sup_{t\in [t_{*}, t_m]  }|\widehat{\sigma}_{j, \lambda}^2(t)-\sigma^2(t)|
	=O_p(h+\sqrt{\log (1/\eta)\eta/h})
	\label{P.2}
	\end{align}
	%where $\hat{\beta}_{jr, \lambda}(t)$ is the $r$th component of $\hat{\bs{\beta}}_{j, \lambda}(t)$, for $r=1, \ldots, d$.
\end{theorem}

To the best of our knowledge, the uniform convergence rate of one-sided kernel smoother with correlated errors has not been thoroughly investigated in the literature. In Theorem \ref{thm1}, we establish the uniform consistency results of the proposed estimators given the time series are $\rho$-mixing. In the right-hand side of \eqref{P.1}, $O(h)$ is a bound for bias while $O_p(\sqrt{\log (1/\eta)\eta/h })$ is a bound for the maximum level of variation. Hence, to make the estimators uniformly consistent, $h$ (equivalently $\lambda$), that is similar to the bandwidth in classical nonparametric regression, is required to satisfy $h\to 0$ and $h/\{\eta\log (1/\eta)\}\to \infty$. Compared to the results of local polynomial smoothers in nonparametric regression, the vanishing rate of the bias in our estimator is $h$ rather than $h^2$, which is due to the use of one-sided kernel.

%{\color{red} While testing the locally misspecified alternatives is beyond the scope of this paper, we assume that for $\delta_j(t)$ that $\frac{1}{\sqrt{-\eta\log \lambda}}\delta_j(t)>\delta_0$ for $j \in \mathcal{O}_t^{(1)}$ and $\frac{1}{\sqrt{-\eta\log\lambda}}\delta_j(t)<-\delta_0$ for $j \in \mathcal{O}_t^{(2)}$, where $\delta_0$ is a positive constant, uniformly in $t$?}

Next theorem provides the uniform convergence rates of $\tilde{\bm\beta}_{\lambda}(t)$ and $\tilde{\sigma}_{\lambda}^2(t)$.
\begin{theorem}\label{thm2}
	Under Assumptions \ref{A1}-\ref{A5}, we have
	\begin{align}\label{eq:supmed}
	\sup_{t\in  [t_{*}, t_m] } \| \tilde{\bm\beta}_{\lambda}(t)-\bm\beta(t) \|
	&=O_p\left(\sqrt{\log (1/\eta)\eta/(hp^{ \zeta }) }\right), \\
	\sup_{t\in [t_{*}, t_m] } |\tilde{\sigma}^2_{\lambda}(t)-\sigma^2(t)|
	&= O_p\left(\sqrt{\log (1/\eta)\eta/(hp^{ \zeta }) }\right),
	\end{align}
where $\zeta$ satisfies that $\frac{\sqrt{h/\eta}/\log(1/\eta)}{1/\eta} \times p^{\zeta} =o(1)$ and $\zeta \le \xi$ with $\xi$ appeared in Assumption~\ref{A2}.
\end{theorem}

The incremental rate $p^{-\zeta }$ due to information fusion is bounded by both $ \frac{\sqrt{h/\eta}/\log(1/\eta) }{1/\eta}$ and $p^{-\xi}$, where $\frac{ \sqrt{ (h/\eta)^{1/2}/\log(1/\eta) } }{ \sqrt{1/\eta} } $ is the Berry-Esseen bound of the $\sqrt{h/\eta}\{\hat{\bbe}_{j, \lambda}(t)-\bbe(t)\}$ under $\rho$-mixing assumption as indicated in Lemma~\ref{BE} of the Supplement and $p^{\xi}$ represents the effective block number. As shown in Theorem \ref{thm1}, as long as $h=(1/\eta)^{-c},\ c>1/3$ so that the bias in $\hat{\bm\beta}_{j,\lambda}(t)$ is negligible and $\zeta <\min(2/3, \xi)$ by the condition, we observe that $\tilde{\bm\beta}_{\lambda}(t)$ has a faster rate of convergence than does $\hat{\bm\beta}_{j,\lambda}(t)$, which suggests a significant efficiency again due to the fusion of information across streams.
This uniform convergence result is particularly helpful for justifying the role of $\tilde{\bm\beta}_{\lambda}(t)$ and  $\tilde{\sigma}^2_{\lambda}(t)$ in the dynamic screening procedure described in Section~\ref{Sec-2.2}. Though uniform convergence results for kernel estimation with dependent data were discussed in the literature, such as \cite{Hansen08} and \cite{vogt2017classification}, technical arguments for this theorem are highly non-trivial and may be interesting in their own rights because our quantile-based estimator $\tilde{\bm\beta}_{\lambda}(t)$ is not a linear statistic.

Next result shows that $\widehat{\lambda}(t_m)$ is asymptotically optimal in the sense that it minimizes the averaged predictive squared error.
\begin{theorem}\label{thm3}
Under Assumptions \ref{A1}-\ref{A5}, provided that the number of outlying datastreams $\mathcal{I}_{t_m}$ is negligible, as $\eta \to 0$, we have
	\[
	\frac{\mbox{\rm APSE}_{\widehat{\lambda}(t_m)}(t_m)}{\inf_{\lambda\in(0,1)}\mbox{\rm APSE}_{\lambda}(t_m)}\rightarrow 1
	\]
in probability, where $\widehat{\lambda}(t_m)$ is obtained by restricting $h=-1/\log \lambda\in [C(1/\eta)^{-1/2+\delta}, \infty)$ in \eqref{lhe} for some constants $C>0$ and $\delta>0$.
\end{theorem}
Theorem~\ref{thm3} is derived under the assumption that $\mathcal{I}_{t_m}$ is not contaminated by many alternative datastreams. The choice of $\mathcal{I}_{t_m}$ proposed in Section~\ref{Sec-2.3} ensures that the proposed method is able to deliver satisfactory performance in both simulations and the real-data example.

Our main theoretical result on the asymptotic validity of the DTS method for both FDP and FDR control is given by the next theorem. We need an additional condition on the change magnitude.

\begin{assumption}\label{A6}
As $\eta\to 0$, $\psi_m \to \infty$, where $\psi_m=|\mC_{\mu}(t_m)|$, $\mathcal{C}_{\mu}(t_m)=\{j \in \mathcal{O}_{t_m}: |\gamma_{j}(t_m)|/\nu_m\to\infty,(m-\tau_j)h\to\infty\}$ and
$\nu_m=\sqrt{\log(p/\eta)/(p^{\zeta}h/\eta)}$.
\end{assumption}

\remark 1
Assumption \ref{A6} is a technical condition for establishing the FDP control of DTS. The $\nu_m$ represents the convergence rate of $\tilde{\bm\beta}_{\lambda}(t)$ and $\tilde{\sigma}^2_{\lambda}(t)$ as discussed in Theorem \ref{thm2}, and the implication of this assumption is that the number of outlying streams with identifiable signal strengths is not too small as $\eta\to 0$ and $p\to\infty$. This seems to be a necessary condition for FDP control under the sparse scenario, say $\Card(\mathcal{O}_{t_m})=o(p)$. For example, in the context of multiple testing, \cite{liu2014phase} showed that even with the true $p$-values, no method is able to control FDP with a high probability if the number of true alternatives is fixed as the number of hypothesis tests goes to infinity. To see this clearer, notice that the key step is to show the validity of (\ref{fdpl}) in which the convergence of empirical sum such like $\sum_{j \notin \mathcal{O}_{t_m}}\bI(\widehat{\gamma}_{j,\lambda}(t_m) \ge u)$ is needed. When $u$ is extremely large, the number of nonzero terms in the summation would be finite and consequently the convergence would fail. The condition that $\psi_m\to\infty$ helps to rule out such cases.

\begin{theorem}\label{thm4}
Suppose Assumptions \ref{A1}-\ref{A6} hold.  For any $\alpha\in(0,1)$, the $\mathrm{FDP}$ of the $\mathrm{DTS}$ method satisfies
${\rm FDP}(t_m)\leq\alpha+o_p(1)$ uniformly at $t_m$. It follows that $\mathop{\lim\sup}_{(1/\eta,p)\to\infty}{\rm FDR}(t_m)\leq \alpha$ uniformly at $t_m$.
\end{theorem}

This theorem shows that the DTS procedure can control the FDR level uniformly at $t_m$. Numerical study shows that our data-driven FDR control approach works well in finite-sample cases and thus greatly facilitates our screening procedure. The next result shows that our DTS is capable of not only controlling the FDR level, but also maintaining the property that all of the outlying streams with certain signal magnitudes can be identified provided sufficient observations are collected after change occurs.

\begin{corollary}
Suppose Assumptions \ref{A1}-\ref{A6} hold. For any $j \in \mathcal{O}_{t_m}$ satisfying $|\gamma_{j}(t_m)|/\nu_m\to\infty$ and $(m-\tau_j)h\to\infty$, we have $\Pr(j\in\widehat{\mathcal{O}}_{t_m})\to 1$ uniformly at $t_m$.
\end{corollary}

Finally, we provide point-wise consistency results for estimating $\pi_{r, t_m}$.

\begin{proposition}\label{prop1}
Under assumptions \ref{A1}-\ref{A5}, and given that $\|\tilde{\bbe}_{\lambda}(t_{m-1}) - \bbe(t_{m-1})\| = o_p(1)$, where $\tilde{\bbe}_{\lambda}(t_{m-1}) $ is the estimated coefficient at $t_{m-1}$ that incorporates the random quantile $\hat{\pi}_{r, t_{m-1}} $, we have
	\begin{align*}
	\frac{1}{p}&\sum_{j=1}^{p}\left\{\ID\{\hat{\beta}_{jr, \lambda}(t_m)>\tilde{\beta}_{r, \lambda}(t_{m-1})\}-\ID\{\hat{\beta}_{jr, \lambda}(t_m)<\tilde{\beta}_{r, \lambda}(t_{m-1})\}\right\}\\
& = \pi_{r, t_m}^{+} -\pi_{r, t_m}^{-}+o_p(1).
	\end{align*}
\end{proposition}

\section{Simulation}\label{Sec-4}

\subsection{Simulation setup}

To demonstrate the finite sample performance of the proposed method, we consider model (\ref{model1}) with time dependent $\mathbf{X}_{ij}  = \{1, X_{ij}\}^{\top}$ covariates, where $X_{ij}$ is generated from the mean zero Gaussian process with $\text{cov}\left\{ X_{j_1}(t_{i_1}),X_{j_2}(t_{i_2} )  \right\} = 0.8^{| t_{i_1}-t_{i_2} |}$. To mimic the real world scenario, our data generating process incorporates not only between-stream dependence but also temporal correlation introduced by the noise variable. In doing so, for each stream $j$, we first generate $ \varepsilon_j := \left(\varepsilon_{1j}, \ldots, \varepsilon_{Nj} \right)^\top$ from a mean zero Gaussian process with autocorrelation $\rho_{\text{Tempo}} \in\{0, 0.5\}$. Then, we multiply the stacked noise matrix $( \varepsilon_1, \ldots, \varepsilon_p )^\top$ with a diagonal block structured correlation matrix $\Sigma_{\text{Block}}$ of block size $n_{\text{Block}} = 200$, and the within block correlation is set to be $\rho_{\text{Block}} \in \{ 0, 0.5\}$. The number of streams $p = 800$. The noise level $\sigma^2(t)\in \{1, 8\}$. Let $t_i= i$ for $i=1, \cdots, N$, and we use the first 300 time points as warm-up period.  Finally, we consider two cases in which the time points $N\in \{ 2400, 4800 \}$.

Next, for any $t\in \{t_1, \ldots, t_N\}$, we generate the coefficient $\bbe(t) = \left( 1,  \beta(t) \right)^{\top}$ through
\begin{align}\label{eq:betasim}
\beta(t) = \begin{cases}
\frac{\sin\left\{ (14s)^{1.5} -14s \right\} \exp\left(7s \right)}{20} +3, & \text{ if }s = \frac{t}{N}\in (0, \frac{1}{2}),\\
\frac{\sin\left\{ (7-14s)^{1.5} - 7+14s \right\} \exp\left(7-7s\right)}{20}+3, & \text{ if } s= \frac{t}{N}\in [\frac{1}{2},1].
\end{cases}
\end{align}
A pictorial illustration of $\beta(t ) $ is given in Figure \ref{fig:choice-lambda}.
 For $t\leq  \frac{N}{2}$, we generate the drift $\bm{\delta}_j(t) = \left( \delta_j(t), 0 \right)^{\top}$ with the following signal lengths and patterns:
\begin{align*}
\delta_{j}(t) =  \begin{cases}
10, & \text{ if } j\in \{1,\ldots, p/10\},\ t\in [ \frac{N}{6}+1, \frac{N}{4}] \cup [ \frac{N}{3}+1, \frac{11N}{24} ],  \\
1, & \text{ if } j\in \{p/10+1,\ldots, p/5\},\ t\in [ \frac{N}{6}+1, \frac{N}{4}] \cup [ \frac{N}{3}+1, \frac{11N}{24} ],  \\
0, & \text{otherwise.}
\end{cases}
\end{align*}
We refer to this period with rather stable change points as ``fixed signal period."
When $t\in [\frac{N}{2}+1,N]$, we consider a more realistic scenario: once a signal occurs, both the signal strength and length change over time. For $j=1,\ldots,p$, assume that the $j$th stream has $\tilde{T}_j$ change points $\tau_{j1}, \cdots, \tau_{j,\tilde{T}_j}$, where $\tilde{T}_j= T_j\mathbf{1}_{T_j\leq 5}$, and $T_j$ follows Poisson distribution with mean 3. Under this data generating process, overall about $1/5$ datastreams contain signals.  Then, to well separate the signals between two adjacent changes points, we generate random change points under the constraint that $|\tau_{jk} - \tau_{j,k+1} | > 200$, for $j=1,\ldots,p$, $k=1,\ldots, \tilde{T}_j-1$. The length of the signals is randomly sampled from $\text{Uniform}[30,80]$. The size of the signal is a nonlinear function of $t$
\begin{align*}
\delta_{j}(t) = \frac{1}{3}\sin \left( \frac{9t}{2N}\pi \right)+\omega_j, \text{ if } j \in \mathcal{O}_t,
\end{align*}
where $\omega_j$ is either 2  or 7 based on a random draw. The smoothing parameter is adaptively chosen from
$\lambda_l = \exp\left( -C_lN^{-0.3} \right), C_l = 0.10 + {l}/{10},l=1,\ldots,10.$
The numerical results presented in this section are evaluated through 200 Monte Carlo replications.

\subsection{Simulation results for dynamic estimation}

\subsubsection{Competing estimation procedures}
We aim to compare the proposed DTS procedure with those reached by ignoring the distribution shift of the process. In the traditional VC model, based on the observed data up to a certain time point $t_m$, the coefficient $\bm\beta(t)$ can be obtained  by minimizing a ``pooled" local loss function
\begin{align*}
& Q_{\text{pool}}(t_m) := \sum_{j=1}^{p}\sum_{i=1}^{m} (y_{ij} - \mathbf{X}_{ij}^{\top}\mathbf{b})^2\lambda^{t_m-t_i},
\end{align*}
and we obtain a naive pooled estimator
\begin{align}\label{poolest}
\displaystyle\widehat{\bbe}_{\lambda, \text{pool}}(t_m) :=\left\{ \sum_{j=1}^{p}\sum_{i=1}^{m}w_i(t_m)\X_{ij}\X_{ij}^{\sT}\right\}^{-1} \sum_{j=1}^{p}\sum_{i=1}^m w_i(t_m)\X_{ij}y_{ij}.
\end{align}
Alternatively, once an estimator $\widehat{\bbe}_{j,\lambda}(t_m)$ is constructed for each stream in the adaptive manner, we may simply take the average as a final estimate
$${\displaystyle
	\widehat{\bbe}_{\lambda, \text{mean}}(t_m) :=  \frac{1}{p}\sum_{j=1}^{p} \widehat{\bbe}_{j,\lambda}(t_m) =\frac{1}{p}\sum_{j=1}^{p}\left\{\sum_{i=1}^{m}w_i(t_m)\X_{ij}\X_{ij}^{\sT}\right\}^{-1}\sum_{i=1}^m w_i(t_m)\X_{ij}y_{ij}.
}$$

\subsubsection{Tuning parameter selection and RMSE comparison}

\begin{figure}[ht]
	\centering
	\includegraphics[width=0.8\linewidth]{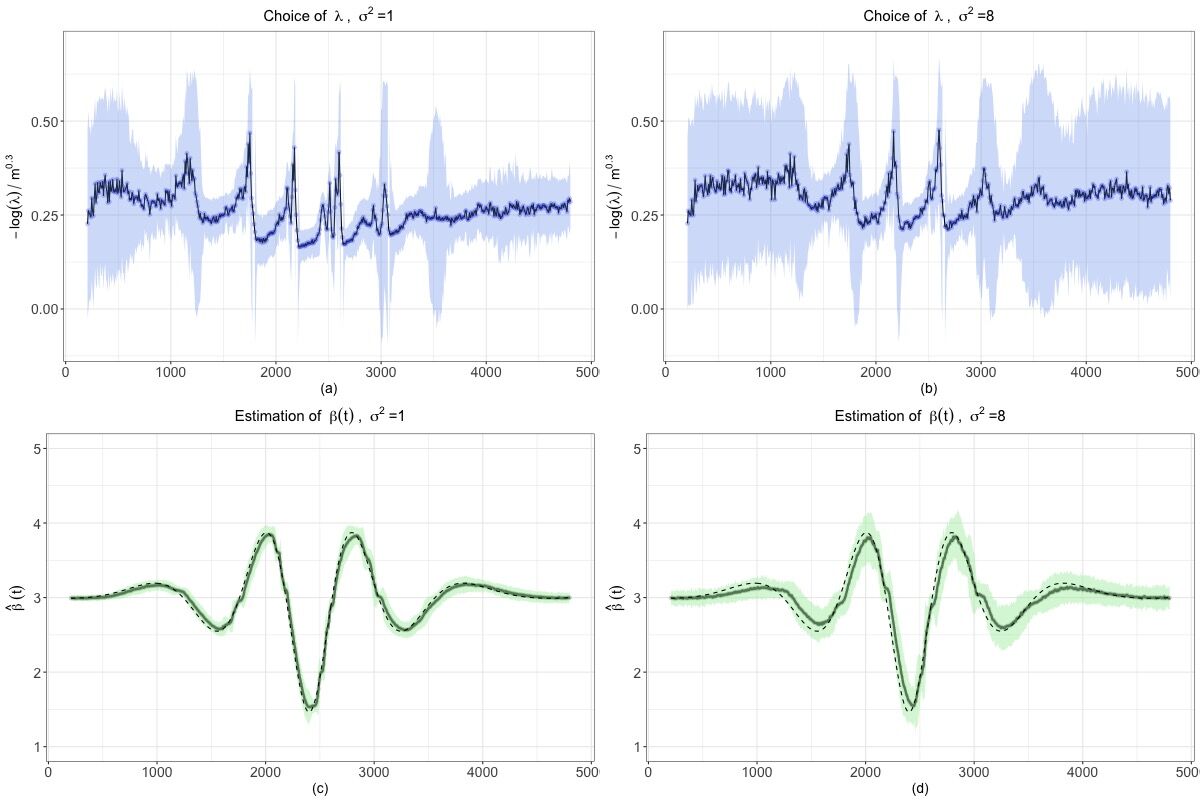}
	\caption{\small For independent streams without temporal correlation $(N,p) = (4800, 800)$ and $\sigma^2(t) \in  \{1,8\}$: (a)-(b): the mean of the adaptively selected $\lambda(t)$ (solid curve) along with the 95\% confidence band (shaded area); (c)-(d): the mean of the proposed estimator $\tilde{\bbe}_{\hat\lambda(t)}(t) $ (solid curve) along with the 95\% confidence band (shaded area). }
	\label{fig:choice-lambda}
\end{figure}

As we discussed earlier, the choice of the tuning parameter $\lambda$ is critical for any smoothing-based procedures. To demonstrate the benefit of the proposed method when the $\lambda$ is adaptively chosen, we compare the performance of $\tilde{\bbe}_{\hat\lambda(t)}(t) $ with $\widehat{\bbe}_{\lambda, \text{pool}}(t) $, $\widehat{\bbe}_{\lambda, \text{mean}}(t)$ and $\tilde{\bbe}_{\lambda}(t) $, while $\lambda$ is fixed across all time points. The average of 200 estimators $\tilde{\bbe}_{\hat\lambda(t)}(t) $ along with the 95\% confidence band is presented in Figure~\ref{fig:choice-lambda}(c)-(d) when $(N,p)=(4800, 800)$ and $\sigma^2(t)$ is either $1$ or $8$ for independent streams (i.e., $\rho_{\text{Tempo}}=0$). The corresponding results for an adaptively selected $\lambda(t)$ are provided in Figure~\ref{fig:choice-lambda}(a)-(b). We observe that the proposed one-step-prediction-based method is capable of adapting the smoothness, and, at the same time, provides accurate estimates of the true underlying varying coefficients $\bs\beta(t)$.

As an implication of the result in Figure~\ref{fig:choice-lambda}, $\lambda_3 = \exp(-0.3N^{-0.3})$ seems to be a good choice for a fixed tuning parameter, which minimizes the average predictive square error for the majority of the time points. Therefore, we proceed with reporting finite sample performances of the proposed estimator in comparison with $\widehat{\bbe}_{\lambda, \text{pool}}(t) $, $\widehat{\bbe}_{\lambda, \text{mean}}(t)$ and $\tilde{\bbe}_{\lambda}(t)$ when $\lambda =\lambda_3$. The results, shown in Table \ref{table:E1}, are the averaged root-mean squared errors (RMSE) of the considered estimators over Monte Carlo samples, defined as
\begin{align*}
\frac{1}{200}\sum_{iter=1}^{200} \left\{ \frac{1}{N}\sum_{i=1}^{N} \|\hat{\bbe}_{iter}(t_i) - \bbe(t_i)\|^2\right\}^{1/2}.
\end{align*}

\begin{table}
	\captionsetup{width=.95 \linewidth}
	\caption{\label{table:E1} \small Root-mean squared errors of the estimated ${\bbe}(t)$}. %The results are based on the the description in Section 1.1. }
	\centering
	\begin{tabular}{cccccc} \hline\hline
	 &&	 $\tilde{\bbe}_{\lambda_{\text{apt}}}(t) $  & $\tilde{\bbe}_{\lambda}(t) $ &  $\widehat{\bbe}_{\lambda, \text{pool}}(t) $ & $\widehat{\bbe}_{\lambda, \text{mean}}(t)$  \\[2ex]
	 	 & \multicolumn{5}{c}{Independent streams without temporal correlation}  \\[2ex]
$N=2400$	&$\sigma^2(t) = 1$ &$ 0.042_{( 0.007 )}$ & $ 0.027_{( 0.008 )}$ & $ 0.438_{( 0.012 )}$ & $ 0.440_{( 0.012 )}$ \\
&	$\sigma^2(t) = 8$ &$ 0.094_{( 0.004 )}$ & $ 0.072_{( 0.002 )}$ & $ 0.430_{( 0.012 )}$ & $ 0.432_{( 0.012 )}$ \\ [2ex]

	 $N=3600$		&	 $\sigma^2(t) = 1$ &$0.014_{( 0.001 )}$ & $ 0.024_{( 0.001 )}$ & $ 0.357_{( 0.012 )}$ & $ 0.360_{( 0.012 )}$ \\
	 &$\sigma^2 (t)= 8$ &$ 0.056_{( 0.002 )}$ & $ 0.054_{( 0.001 )}$ & $ 0.358_{( 0.012 )}$ & $ 0.358_{( 0.012 )}$   \\[2ex]
	
	 $N=4800$	&	 $\sigma^2(t) = 1$ &$ 0.015 _{( 0.001 )}$ & $ 0.023 _{( 0.001 )}$ & $ 0.331 _{( 0.012 )}$ & $ 0.333 _{( 0.012 )}$ \\
	 						   &	$\sigma^2(t) = 8$ & $ 0.047_{( 0.002 )}$ & $ 0.045 _{( 0.001 )}$ & $ 0.326 _{( 0.012 )}$ & $ 0.327 _{( 0.012 )}$ \\  [2ex]
	
	 & \multicolumn{5}{c}{Dependent streams with $\rho_{\text{Block}} = 0.5$, $\rho_{\text{Tempo}} = 0$}  \\[2ex]
	%	\multicolumn{5}{c}{	$m=2400$, $\lambda_{\text{fix}} = \lambda_3$, $\lambda_{\text{mean}} = \lambda_8$}\\
	
		$N=2400$	&	 $\sigma^2(t) = 1$ & $ 0.043_{( 0.004 )}$ & $ 0.029_{( 0.007 )}$ & $ 0.443_{( 0.012 )}$ & $ 0.444_{( 0.012 )}$ \\
		 &$\sigma^2 (t)= 8$ &	$ 0.136_{( 0.006 )}$ & $ 0.093_{( 0.002 )}$ & $ 0.466_{( 0.012 )}$ & $ 0.460_{( 0.012 )}$ \\ [2ex]
$N=3600$		&	 $\sigma^2(t) = 1$ & $ 0.017_{( 0.001 )}$ & $ 0.031_{( 0.001 )}$ & $ 0.359_{( 0.012 )}$ & $ 0.361_{( 0.012 )}$ \\
		 &$\sigma^2 (t)= 8$ & $ 0.077_{( 0.003 )}$ & $ 0.103_{( 0.003 )}$ & $ 0.372_{( 0.012 )}$ & $ 0.373_{( 0.012 )}$ \\ [2ex]

  $N=4800$	&	 $\sigma^2(t) = 1$ & $ 0.016 _{( 0.002 )}$ & $ 0.023 _{( 0.001 )}$ & $ 0.324 _{( 0.009 )}$ & $ 0.326 _{( 0.009 )}$ \\
	&	 $\sigma^2(t) = 8$ & $ 0.058_{( 0.002 )}$ & $ 0.061 _{( 0.002 )}$ & $ 0.326_{( 0.012 )}$ & $ 0.328 _{( 0.012 )}$ \\  [2ex]
	
		 & \multicolumn{5}{c}{Dependent streams with $\rho_{\text{Block}} = 0.5$, $\rho_{\text{Tempo}} = 0.5$}  \\[2ex]
	$N=2400$		&	 $\sigma^2(t) = 1$ &$ 0.036_{(0.001 )}$ & $ 0.032_{( 0.001 )}$ & $ 0.430_{( 0.010 )}$ & $ 0.432_{( 0.010 )}$ \\
	&$\sigma^2 (t)= 8$& $ 0.146_{( 0.004 )}$ & $ 0.150_{( 0.003 )}$ & $ 0.437_{( 0.014 )}$ & $ 0.438_{( 0.015)}$ \\ [2ex]
	
	$N=3600$	&$\sigma^2(t) = 1$  &$ 0.028_{( 0.001 )}$ & $ 0.029_{( 0.001 )}$ & $ 0.360_{( 0.009 )}$ & $ 0.361_{(0.010 )}$ \\
	&	 $\sigma^2(t) = 8$ & $ 0.085_{( 0.003 )}$ & $ 0.094_{( 0.002 )}$ & $ 0.370_{( 0.012 )}$ & $ 0.369_{( 0.012 )}$ \\ [2ex]
	
	$N=4800$	&	 $\sigma^2(t) = 1$ & $ 0.019 _{( 0.001 )}$ & $ 0.027 _{( 0.001 )}$ & $ 0.330 _{( 0.012 )}$ & $ 0.332 _{( 0.012 )}$ \\
	&	 $\sigma^2(t) = 8$ & $ 0.078 _{( 0.003 )}$ & $ 0.067 _{( 0.002 )}$ & $ 0.329 _{( 0.012 )}$ & $ 0.332 _{( 0.012 )}$ \\
	\hline\hline	
	\end{tabular}
\end{table}

We find that the proposed method $\tilde{\bbe}_{\hat\lambda(t)}(t) $ has the smallest RMSE among the considered cases. The results also bear out the intuition of Theorem \ref{thm2}, suggesting that the RMSE of $\tilde{\bbe}_{\hat\lambda(t)}(t) $ decreases with the number of total time points $m$ and with the level of the noise $\sigma^2(t)$. Interestingly, we find that the difference between $\widehat{\bbe}_{\lambda, \text{pool}}(t) $ and $\widehat{\bbe}_{\lambda, \text{mean}}(t)$ is nearly negligible. This can be understood by writing down the difference between these two estimators at the given time point $t_m$:
\begin{align}\label{eq:sim1}
{ \footnotesize \frac{m}{p}\sum_{j=1}^{p} \left\{ \left( \frac{1}{mp}\sum_{i=1}^{p}\sum_{i=1}^{m}w_i(t_m)\X_{ij}\X_{ij}^{\sT}\right)^{-1} - \left( \frac{1}{m}\sum_{i=1}^{m}w_i(t_m)\X_{ij}\X_{ij}^{\sT}\right)^{-1} \right\}w_i(t_m)\X_{ij}y_{ij}},
\end{align}
that vanishes with a large $m$, as the difference between the two inverse sample covariance matrices converges to zero. Lastly, by comparing the performances between $\tilde{\bbe}_{\lambda}(t)$ and $\widehat{\bbe}_{\lambda, \text{pool}}(t)$ shows that our quantile based procedure is more robust against the presence of outlying datastreams.

\subsection{Simulation results for dynamic testing}

\subsubsection{Competing testing procedures}
We now move on to compare the DTS testing procedure with three others:

First, we compare the proposed DTS with the nonparametric test of \cite{Zheng_1996} customized to the dynamic environment. For the given time point $t_m$, we consider the following test statistics:
\begin{align}\label{eq:MWBenjamini-HochbergNT}
\frac{1}{n(n-1)}\sum_{\substack{t_k\neq t_i,\\ t_i, t_k\in [ t_m-n+1, t_m ]}}K\left(\frac{t_i - t_k}{b_n}\right)\z_{ij}\z_{kj},
\end{align}
where $n$ is a given window size, $b_n$ is the bandwidth, and $K(u) = 0.75(1-u^2)_{+}$ is the Epanechnikov kernel function for simplicity. After constructing the test statistics and calculating the corresponding p-values by normal approximating, we adopt the Benjamini-Hochberg procedure \citep{Benjamini_Hochberg_1995} to adjust for the effect of multiple comparison. Such a method is referred to as moving-window-based nonparametric test (MWNT).  We showcase MWNT with $n=400$ and $b_n\in\{0.03N, 0.05N\}$.

As the between stream correlation structure can affect the asymptotic distribution of the MWNT statistics defined in \eqref{eq:MWBenjamini-HochbergNT}, to present a fair comparison, we employ the following decorrelation strategy under the assumption that the noise variables are $\omega$-dependent (i.e., $\cov(\varepsilon_{i_1j},\varepsilon_{i_2j})=0$ if $t_{i_1}-t_{i_2}>\omega $ for some constant $\omega$. The correlation, $\cov(\varepsilon_{i_1j},\varepsilon_{i_2j})$ for $\{(i_1,i_2):t_{i'}-t_i\leq \omega \}$, can be estimated with the warm-up period observations. At a given time point $t_m$, suppose ${\bf L}{\bf L}^{\top}$ is the Cholesky decomposition of the correlation matrix for $(\varepsilon_{ij},\ldots,\varepsilon_{t_mj})$ with $t_m-t_i\leq \omega $. Then we can transform the data $\tilde{\bf z}_j(t_m)=(\z_{ij},\ldots,\z_{mj})^{\top}$ by ${\bf L}\tilde{\bf z}_j(t_m) $ whose elements can be used instead of $\z_{ij}$ in \eqref{eq:MWBenjamini-HochbergNT}. In our simulation studies, as the temporal correlation decreases exponentially fast as the time interval increases, we set $\omega = 20$.

The second approach we compare with is based on estimating the long-run covariance matrix via  \cite{andrews1991heteroskedasticity}. There, the author proposes a heteroskedasticity and autocorrelation consistent (HAC) estimation of covariance matrices for the estimated coefficients in linear models. In brief, we consider the following test statistics for each stream $j$ at time $t_m$:
\begin{align}\label{eq:test-stat-andrews}
\frac{ \frac{1}{n_{\text{HAC}}}\sum_{ i: t_i \in [ t_m-n_{\text{HAC}}+1,t_m] } \z_{ij}  }{\hat{\text{Sd}}\big( \frac{1}{n_{\text{HAC}}}\sum_{ i: t_i \in [ t_m-n_{\text{HAC}}+1,t_m] } \z_{ij}  \big)} ,
\end{align}
where $\hat{\text{Sd}}\big( \frac{1}{n_{\text{HAC}}}\sum_{ i: t_i \in [ t_m-n_{\text{HAC}}+1,t_m] } \z_{ij}  \big)$ is the estimated standard deviation reported by the \texttt{R} package \texttt{sandwich}, and $n_{\text{HAC}}$ is a user-specific window size and is set to be $110$ in our simulation study. Given this test statistics, we report the testing results based on the Benjamini-Hochberg's linear step up procedure \citep{Benjamini_Hochberg_1995} and the local false discovery rate \citep{efron2004large} as the screening tool.

Lastly, we compare the performance of the DTS testing procedure based on the naive pooled estimator $\hat{\beta}_{\lambda, \text{pool}}(t)$.

In the following sections,  DTS refers to the proposed testing procedure, DTS (Pooled) refers to the  DTS procedure with $\hat{\beta}_{\lambda, \text{pool}}(t_m)$, HAC-LFDR referes to the testing procedure based on \eqref{eq:test-stat-andrews}  \citep{andrews1991heteroskedasticity}  with the local false discovery rate adjustment  procedure \citep{efron2004large}, HAS-BC refers to the procedure based on \eqref{eq:test-stat-andrews} with the Benjamini-Hochberg procedure to adjust for the multiple comparison effect, and MWNT refers to the testing procedure built on  \cite{Zheng_1996}. Since it is difficult to detect if the dependence between data streams exists in reality, we implement the described decorrelation strategies even if the simulated data streams are independent.

\subsubsection{Computational efficiency comparison}

Computational efficiency is a vital concern to screen out the irregular individuals in the massive datastreams, as of which the primary goal is to find out the signals as soon as they occur. To illustrate the benefits of our proposal, we report the average runtime for DTS, HAC- (with LFDR) and MWNT- (with $b_n=0.03N$) based procedures in Table \ref{table:T1}. We also note that the simulations are paralleled on 50 nodes (each node is equipped with  2.5 GHz Intel Xeon 10-core Ivy Bridge processors) via the packages ``doSNOW" (for parallelization) and ``rlecuyer" (for correct parallelization of random numbers).

\begin{table}[ht]
	%	\captionsetup{width=.95 \linewidth}
	\caption{\label{table:T1} \small Computation time (unit: second) of the three testing procedures for dependent datastreams when $(N,p) = (4800, 800)$, $\rho_{\text{Block}} = \rho_{\text{Tempo}} = 0.5$ and $\sigma^2(t) = 1$. }
	\centering
	\begin{tabular}{cccc} \hline\hline
		&$\text{DTS}$ &  HAC & MWNT \\\cline{2-4}
		$N=2400$  & 35.46 &   6931.77&   1893.55  \\
		$N=3600$ & 50.12  &  8691.21 & 2817.22   \\
		$N=4800$ & 68.49 & 12,724.01 & 3644.08  \\\hline\hline
	\end{tabular}
\end{table}

From the results in Table \ref{table:T1}, due to the usage of updating formulae in Section \ref{Sec-2.1}, our DTS procedures have clear advantages over MWNT and the HAC procedures. This advantage also suggests the need of careful designs when using a standard model specification test in a dynamic streaming environment.

\subsubsection{FDR, TPR and the length of delay comparison }

\begin{figure}[ht]
	\centering
	\includegraphics[width=0.9\linewidth]{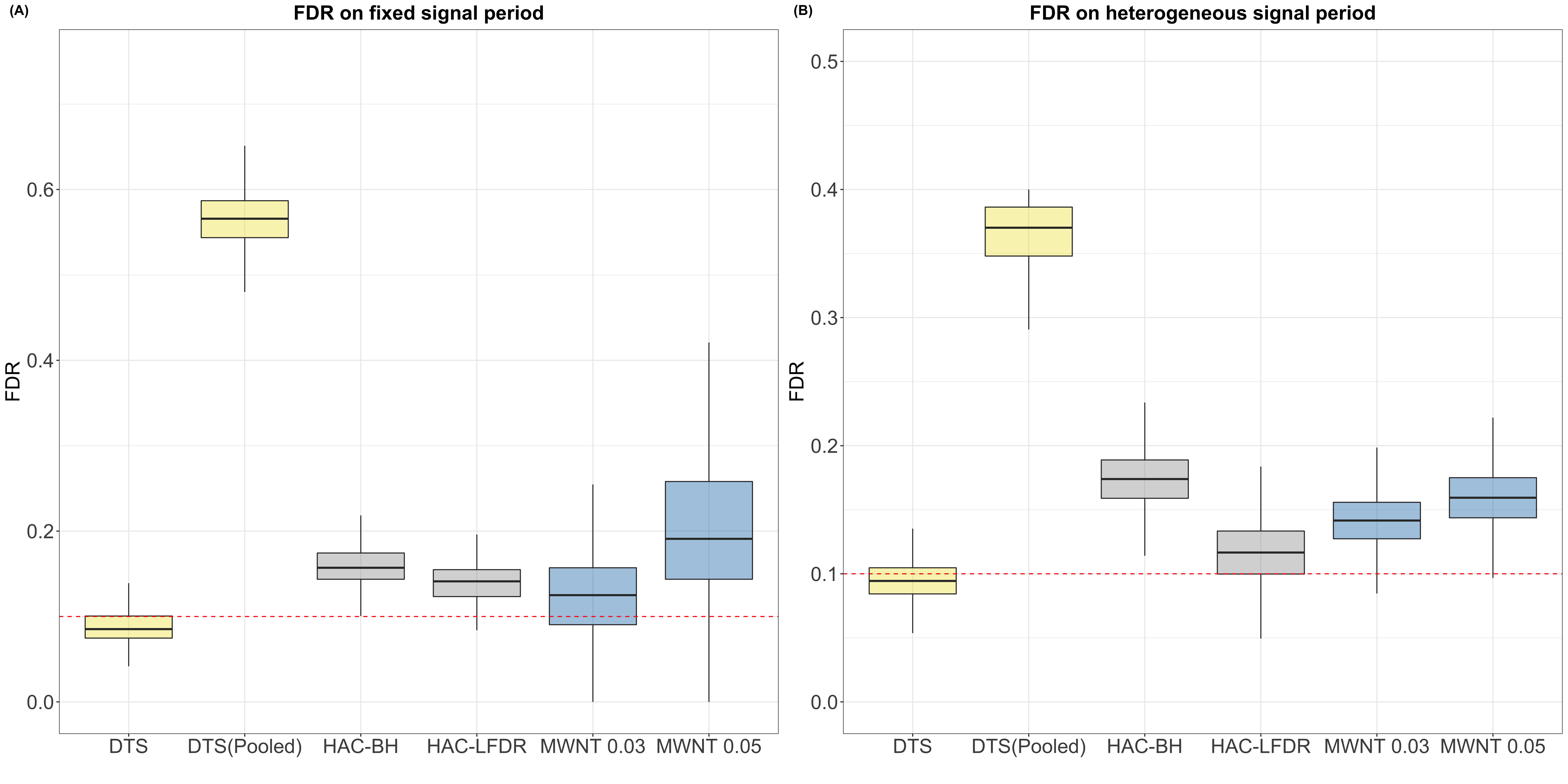}
	\caption{\small For independent streams without temporal correlation, $(N,p) = (4800, 800)$, $\sigma^2(t) = 1$: Figure (A) is the boxplot of the empirical FDR in the fixed signal period; Figure (B) is the boxplot of the empirical FDR in the heterogeneous signal period.}
	\label{fig:fdr}
\end{figure}

\begin{figure}[ht]
	\centering
	\includegraphics[width=0.9\linewidth]{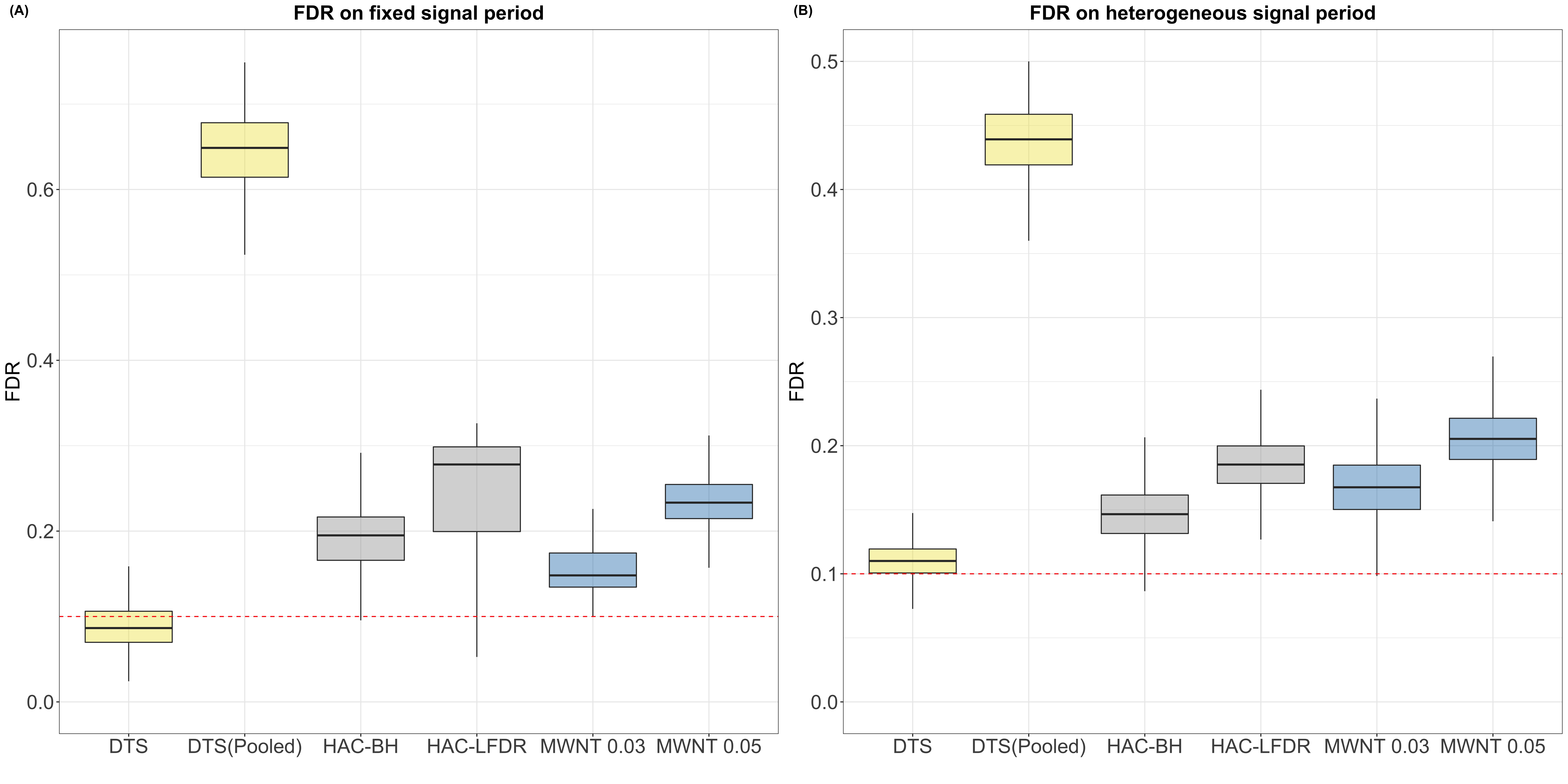}
	\caption{\small For dependent streams $\rho_{\text{Block}}=0.5$ with temporal correlation $\rho_{\text{Tempo}}=0.5$, $(N,p) = (2400 , 800)$, $\sigma^2(t) = 8$: Figure (a) is the boxplot of the empirical FDR in the fixed signal period; Figure (b) is the boxplot of the empirical FDR in the heterogeneous signal period.}
	\label{fig:fdr-dependent}
\end{figure}

\begin{figure}[ht]
	\centering
	\includegraphics[width=0.8\linewidth]{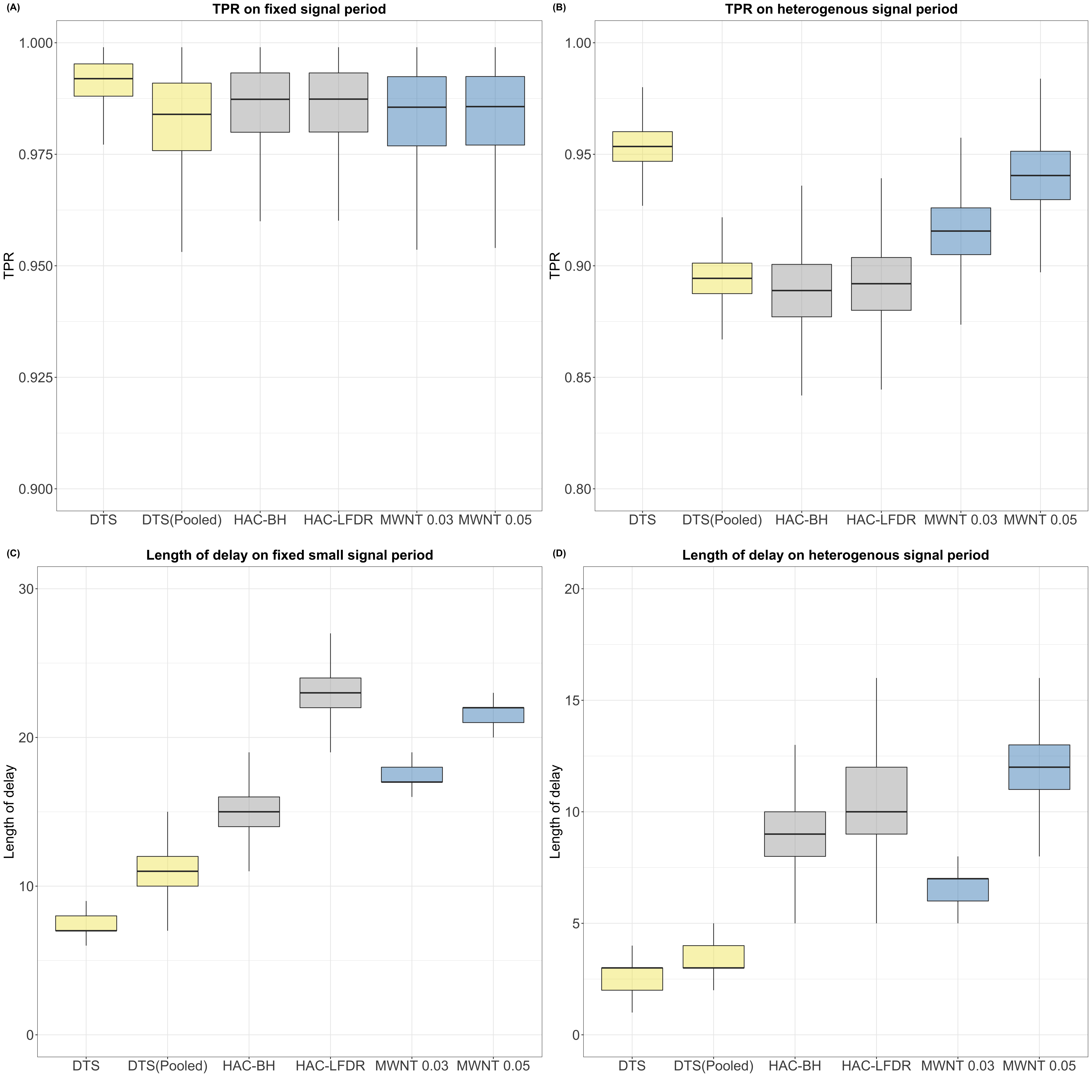}
	\caption{\small For independent streams without temporal correlation, $(N,p) = (4800, 800)$, $\sigma^2(t) = 1$: the boxplots of the medians of TPRs and detection delays in the fixed signal and the heterogeneous periods.}
	\label{fig:delay}
\end{figure}

\begin{figure}[ht]
	\centering
	\includegraphics[width=0.75\linewidth]{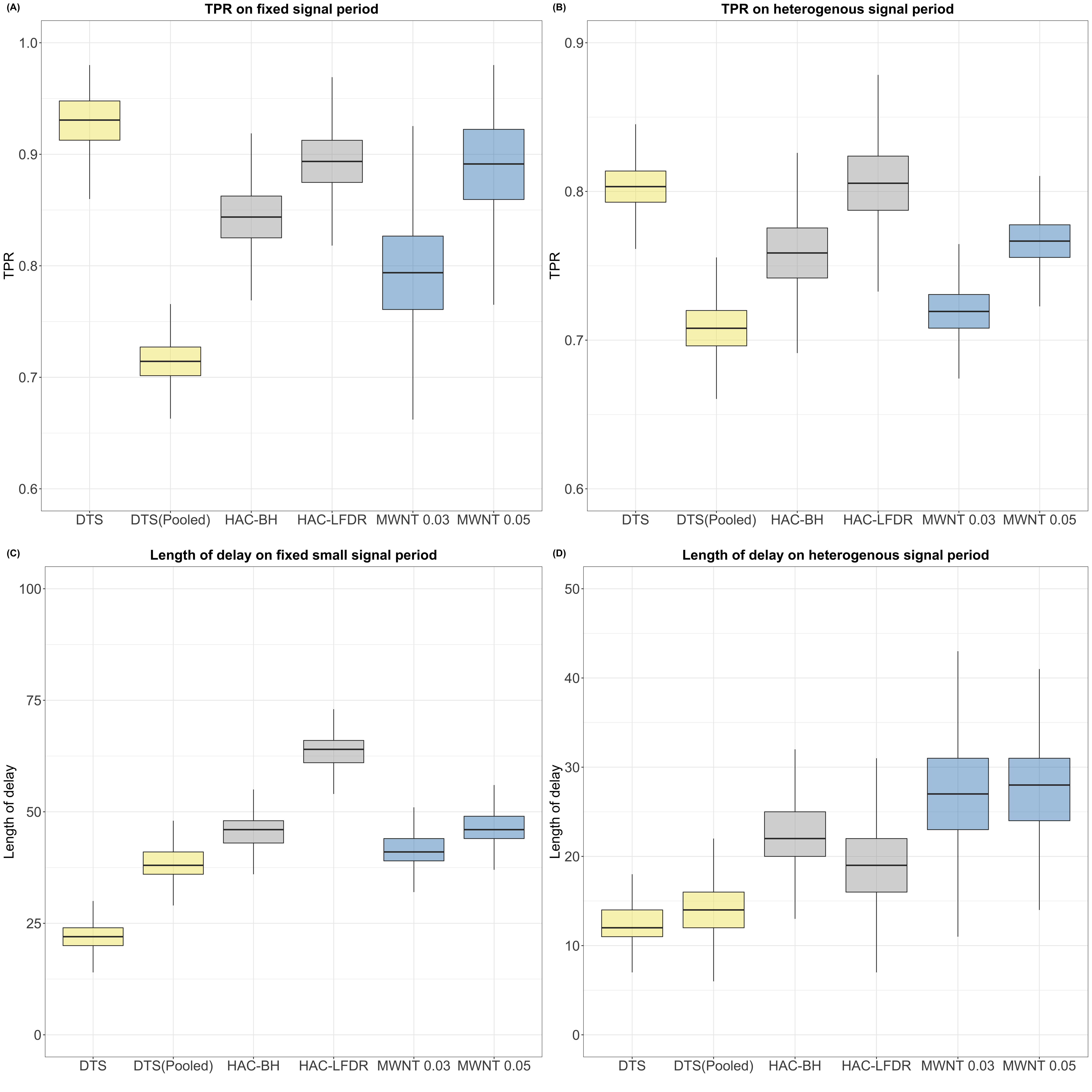}
	\caption{\small The boxplots of the medians of TPRs and detection delays in the fixed signal and the heterogeneous periods for dependent streams $\rho_{\text{Block}}=0.5$ with temporal correlation $\rho_{\text{Tempo}}=0.5$, $(N,p) = (2400, 800)$, $\sigma^2(t) = 8$.}
	\label{fig:delay-dependent}
\end{figure}

In this section, we evaluate the performance of DTS in terms of FDR controls, true positive rate (TPR), and the length of delay. The latter one is defined as the minimum number of steps that a procedure takes to detect a signal within each alternative period. In the literature of sequential change detection (e.g., see \cite{Zou_etal_2015}), this detection delay corresponds to the well-known {\it run-length}.  If there is no such signal of detection, the detection delay is simply set as the length of that specific signal period. We record the medians of detection delays and TPRs amongst all shift periods in each replication. The nominal FDR level is set to be 0.1. To avoid redundancy, in this part, we report the simulation results in two extreme cases: (i) independent datastreams without temporal correlation, $(N,p) = (4800,800)$ and noise level $\sigma^2(t) = 1$, and (ii) dependent datastreams with temporal correlation, $(N,p) = (2400,800)$ and noise level $\sigma^2(t) = 8$. Figure \ref{fig:fdr} and \ref{fig:delay} show  results for independent datastreams, and Figure \ref{fig:fdr-dependent} and \ref{fig:delay-dependent} show results for dependent datastreams.


For the false discovery rate control, from the results in Figures \ref{fig:fdr} and \ref{fig:fdr-dependent}, we observe that both DTS and, to a lesser extent, MWNT- and HAC- based procedures control FDR at desired levels within acceptable ranges when the datastreams are independent. The Pooled-DTS procedure, built on the inconsistent estimate of $\bbe(t)$, fails to control the FDR at the nominal level. When datastreams are correlated, the decorrelation-based testing procedures (HAC-BH, HAC-LFDR and MWNT) tend to have higher FDR than the independent case. In a sharp contrast, the new DTS procedure that utilizes the empirical distribution to approximate the number of false positives is able to deliver very accurate FDR control irrespective of the correlation structures and signal patterns.

In terms of signal detection comparison displayed in Figure \ref{fig:delay}-\ref{fig:delay-dependent},  not surprisingly, since the pooled-DTS does not take the outliers into account and $\bs\beta(t)$ is poorly estimated, it yields low detection power.
Our DTS method outperforms the other competitors by a significant margin in both scenarios from the viewpoint of detection delay. It also delivers satisfactory performance in terms of TPR.
Compared to DTS, MWNT- and HAC-based procedures show comparative detection power due to the signal accumulation, but the long detection delay seems to be unavoidable. Based on the same reasoning, MWNT shows higher detection power with $b_n = 0.05N$, although longer bandwidth yields longer detection delay. In sum, the DTS addresses the need to dynamically detect the streaming pattern and provides more robust performance than the existing methods from the viewpoints of detection delay and power.

\section{Analysis of Intern Health Study Data}\label{Sec_IHS}

The Intern Health Study (IHS) is an on-going mobile health cohort study that enrolls more than 3,000 interns annually. The long-term goal of the IHS is to elucidate the pathophysiological architecture underlying depression to facilitate the development of improved treatments. One important goal of IHS is to identify subjects with short-term risks for mood changes whenever new data from the mobile app ``MyDataHelps" are updated on a daily basis.

In this case study, for illustration purposes, we have restricted our attention to the 2018 IHS cohort with $1,565$ subjects enrolled from July to December. During the study period, the study tracks medical interns using phones and wearables. The outcomes are daily self-reported mood valence (measured through a one-question survey; one of two cardinal symptoms of depression, \cite{lowe2005detecting}). Participants are prompted to enter their daily mood rated from 1-10 every day at a user-specified time between 5 pm and 10 pm.  The time-varying covariates are daily steps prior to the survey (as a proxy for activity) and daily sleep duration that ended in the same day. Both covariates are important potential predictors of mood \citep{kalmbach2018effects}. Our data set thus consists of data from  $p=1,565$ subjects over $m=182$ days. We treat the first $40$ days as the warm-up phase where data are used to initialize the varying coefficient $\bm{\beta}{(\cdot)}$ estimate. This warm-up phase can be viewed as study baseline, because during this period work and life routines are being established and subjects are usually  not suffered from sustained stress.

%\begin{figure}[ht]
%	\centering
%	\includegraphics[width=0.7\linewidth]{sleep-step.jpeg}
%	\caption{ The estimated main time-varying effects for (left) step counts (cubic root scale) and (right) sleep hours (square root scale).% { Let's also show this at the linear predictor scale.}
%	}
%	\label{fig:ihs_beta}
%\end{figure}

%\begin{figure}[ht]
%	\centering
%	\includegraphics[width=0.75\linewidth]{detected_times_two_random_people.pdf}
%	\caption{ The estimated $\mathbf{X}_{ij}^\top\bm{\delta}_{j}(t_i)$ for two { random} subjects{.} The red horizontal {pluses} indicate the days with detected deviation from the population model; {both} are online estimates/decisions.}
%	\label{fig:ihs_two_subjects}
%\end{figure}

\begin{figure}[ht]
	\centering
	\includegraphics[width=0.75\linewidth]{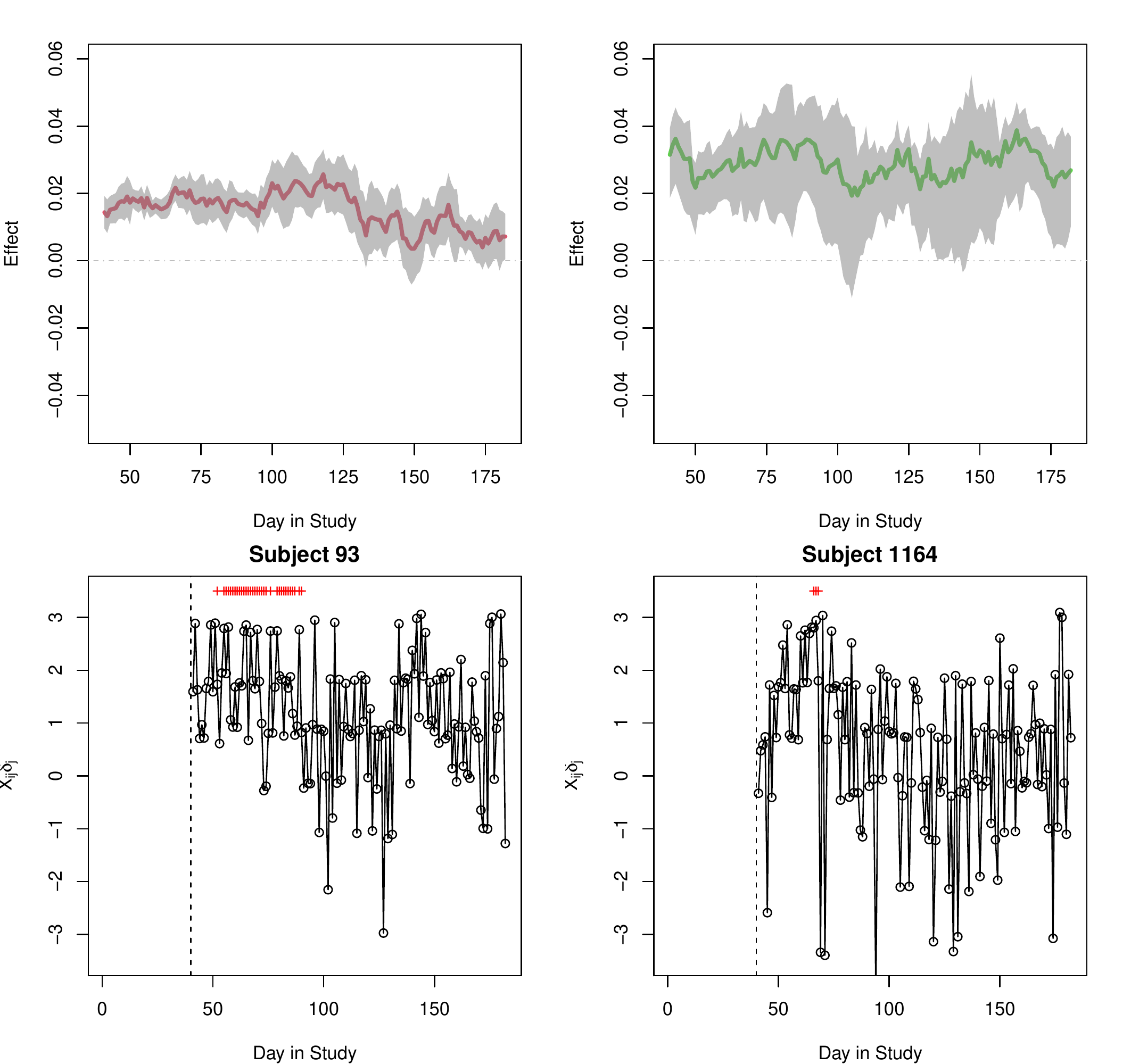}
	\caption{Upper panels: the estimated main time-varying effects for (left) step counts (cubic root scale) and (right) sleep hours (square root scale; Lower panels: the estimated $\mathbf{X}_{ij}^\top\bm{\delta}_{j}(t_i)$ for two { random} subjects{.} The red horizontal {pluses} indicate the days with detected deviation from the population model; {both} are online estimates/decisions.}
	\label{fig:data}
\end{figure}

The upper panels of Figure~\ref{fig:data} show the \textit{online} time-varying effect estimates of daily step counts (cubic root transformed; left) and daily sleep hours (square root transformed; right) upon the mood score along with $95\%$ pointwise confidence bands. In other words, in each time point, we only use the data point collected prior to this time point. {Although} the actual {magnitudes} of the two estimated effect curves depend on the scale of the predictors, we obtain prominent positive effect over time for both the step and sleep predictors indicating their dynamic {and} positive effects upon daily mood uniformly over time. The proposed DTS method also detects periods when an individual's mood trajectory as a function of time, sleep hours and step counts cannot be described by a null population model. In the lower panels of Figure \ref{fig:data}, we illustrate for two randomly chosen subjects the time points {when} we detected such deviations from a population dynamic model  (red pluses at the top). Relative to the subject on the right with mood scores oscillating around values predicted by a population mean model, the subject on the left has mood scores that are too high to be well characterized by the population model. The estimated coefficients indicate longer periods of extra effects of sleep and activity. Our results show that individuals have distinct timings and duration when the joint effect of sleep hours and step counts change the mood to a different degree than others in the population, highlighting the timings to intervene upon sleep and activities, for example, through push notifications via their mobile {phones}. %Incorporation of additional {predictors} such as circadian phase may change the scientific findings that will be reported { elsewhere in the final complete analysis of the study results.}

\section{Concluding remarks}\label{Sec-5}
We conclude this article with three remarks. Firstly, although the local linear kernel estimator has certain advantages over the local constant kernel estimator, as shown in the literature \citep{Fan_1999}, our simulation results show that these two estimators yield similar performance in the dynamic screening. Thus, the local constant kernel procedure (\ref{lf}) is chosen for simplicity. Systematic study of local polynomial smoothers in the present problem warrants future research.

Secondly, one important issue with varying coefficient models is how to incorporate the within-subject correlation structure into the estimation or testing procedure. This issue has been investigated, and the methodology has been well established, especially for longitudinal data analysis; e.g., see \cite{Sun_etal_2007}. For estimation, Theorem \ref{thm2} shows that the consistency of the proposed estimators is valid under quite general correlation assumptions. Though it has been shown that an estimator can be improved by incorporating the within-subject correlation into the estimation procedure \citep{Fan_etal_2007}, such an improved procedure would generally require iterative steps (at least two steps) and the corresponding estimators do not permit recursive calculation. Although it may be computationally feasible to perform a complicated estimation for fixed longitudinal data, fast implementation is likely to be our first priority in massive streaming cases. Certainly, it is of interest to see how the correlation structure can be accommodated into our dynamic tracking procedure.

Finally, in some applications, we may also consider the following individual-specific change-point model
\begin{align}\label{model2}
y_{ij}{=}\left\{\hspace{-0.1cm}
\begin{array}{ll} \X_{ij}^{\sT}\bbe_j(t_i)+\sigma_j(t_i)\varepsilon_{ij}, & {\mathrm {for\ }} \quad t_{i}\in(0,\tau_j],\\[0.1cm]
\X_{ij}^{\sT}\{\bbe_j(t_i)+\bm\delta_j(t_i)\}+\sigma_j(t_i)\varepsilon_{ij}, & {\mathrm {for\ }} \quad t_{i}>\tau_j.
\end{array}\right.
\end{align}
That is, we assume that different datastreams have different coefficient functions $\bbe_j(\cdot)$. The estimation procedure given in Equations (\ref{betare})-(\ref{siup}) is still applicable for this model. However, to make the screening procedure effective, we need to impose additional conditions on $\bm\delta_j(t_i)$, say $\bm\delta_j(t_i)$ is discontinuous at $\tau_j$;  otherwise the change pattern cannot be identified since we are using the nonparametric kernel smoothing approaches. In this way, the screening task can be reframed into an on-line ``jump" detection problem which is well investigated in a non-sequential setting. For jump detection, most existing approaches start with a diagnostic statistic
computed from observations in a local neighborhood of a given point, such as the difference between a right- and a left-sided kernel smoother. Then, a
large value of the diagnostic statistic would indicate a potential jump near the given point. See \cite{Loader:1996ef} and \cite{Gregoire:2002eo} for example. There is a need to investigate how to adapt those methods to the present on-line environment.

%missing data are common in a multiple datastreams application. An implicit assumption made here is that the fraction of missing values is trivial. In such cases, we may simply remove the unobserved individuals from the estimator (\ref{b0e}).  The suggested weighting scheme in (\ref{lce}) is expected to be particularly robust because it can accommodate unequally spaced design points by $\lambda^{t_m-t_i}$.

%\section{Acknowledgments}
%The authors would like to thank Peng Liao, Srijan Sen, Perter Song and Ambuj Tewari for helpful discussions and useful comments, and the Associate Editor and anonymous referees for their constructive comments.
\vspace{0.5cm}
{
\bibliographystyle{apalike}
 \baselineskip 10pt
\bibliography{reference_DTS}

}

\end{document}